\documentclass[hyperref]{aa}  

\usepackage[english]{babel}
\usepackage{natbib}
\usepackage[titletoc,title]{appendix}
\usepackage{graphicx}
\usepackage{epstopdf}
\usepackage{subcaption}
\usepackage{float}
\usepackage{txfonts}
\usepackage{url,color,geometry, siunitx}
\usepackage{amssymb, amsmath}
\usepackage{mathtools}
\usepackage{wasysym}
\usepackage{longtable}
\DeclareGraphicsExtensions{.pdf, .pdf}

\begin{document}

   \title{The adventure of carbon stars \thanks{Based on observations made with ESO telescopes at La Silla Paranal Observatory under program IDs: $090$.D-$0410$, $086$.D-$899$, $187$.D-$0924$, $081$.D-$0021$, $086$.D-$0899$}}

 \subtitle{Observations and modelling of a set of C-rich AGB stars}

   \author{G. Rau \inst{1}
          \and
          J. Hron \inst{1}   
          \and
          C. Paladini \inst{2}                 
          \and
          B. Aringer \inst{3}
          \and
          K. Eriksson \inst{4}
          \and
          P. Marigo \inst{3} 
          \and         
          W. Nowotny \inst{1}
          \and
          R. Grellmann \inst{5}}

   \institute{University of Vienna, Department of Astrophysics, T\"urkenschanzstrasse 17, A-1180 Vienna\\
              \email{gioia.rau@univie.ac.at}
         \and
             Institut d’Astronomie et d’Astrophysique, Universite' libre de Bruxelles, Boulevard du Triomphe CP $226$, B-1050 Bruxelles, Belgium
         \and
             Astronomical Observatory of Padova -- INAF, Vicolo dell’Osservatorio 5, I-35122 Padova, Italy
         \and
             Department of Physics and Astronomy, Division of Astronomy and Space Physics, Uppsala University, Box 516, 75120 Uppsala, Sweden
          \and
             Physikalisches Institut der Universit\"at zu K\"oln, Z\"ulpicher Str. 77, 50397 K\"oln - Germany 
             }

   \date{Received July 18, 2016; accepted January 16, 2017}

 
  \abstract
  {Modelling stellar atmospheres is a complex, intriguing task in modern astronomy. A systematic comparison of models with multi-technique observations is the only efficient way to constrain the models.}
  {We intend to perform self-consistent modelling of the atmospheres of six carbon-rich AGB stars: \object{R Lep}, \object{R Vol}, \object{Y Pav}, \object{AQ Sgr}, \object{U Hya} and \object{X TrA}, with the aim of enlarging the knowledge of the dynamic processes occurring in their atmospheres.}
   {We used VLTI/MIDI interferometric observations, in combination with spectro-photometric data, and compared them with self-consistent dynamic models atmospheres.}
   {We found that the models can reproduce SED data well at wavelengths longward of $1~\mu$m, and the interferometric observations between $8~\mu$m and $10~\mu$m. Discrepancies observed at wavelengths shorter than $1~\mu$m in the SED, and longwards of $10~\mu$m in the visibilities, could be due to a combination of data- and model-related effects. The models best fitting the Miras are significantly extended, and have a prominent shell-like structure. On the contrary, the models best fitting the non-Miras are more compact, showing lower average mass-loss. The mass loss is of episodic or multi-periodic nature but causes the visual amplitudes to be notably larger than the observed ones.	 
A number of stellar parameters were derived from the model fitting: $T_\text{eff}$, $L_\text{bol}$, $M$, $C/O$, $\dot{M}$. Our findings agree well with literature values within the uncertainties. $T_\text{eff}$ and $L_\text{bol}$ are also in good agreement with the temperature derived from the angular diameter $\theta_\text{(V-K)}$ and the bolometric luminosity from the SED fitting $L_\text{bol}$, except for AQ~Sgr. The possible reasons are discussed in the text. Finally, $\theta_\text{Ross}$ and $\theta_\text{(V-K)}$ agree with each other better for the Miras targets than for the non-Miras, which is probably connected to the episodic nature of the latter models.
We also located the stars in the H-R diagram, comparing them with evolutionary tracks. We found that the main derived properties ($L$, $T_\text{eff}$, $C/O$ ratios and stellar masses) from the model fitting are in good agreement with TP-AGB evolutionary calculations for carbon stars carried out with the COLIBRI code.}  
   {}


\authorrunning{Rau et al.}
\titlerunning{The adventure of carbon stars - Observations and modelling of a set of C-rich AGB stars}
\maketitle

\section{Introduction}

Stars less massive than $\sim8$~M$_{\odot}$ and more massive than $~0.8$~M$_{\odot}$, after moving from the Main Sequence through the Red Giant Phase and past the Horizontal Branch, will spend part of their life on the Asymptotic Giant Branch.

At the beginning of the AGB, the stars are characterized by a C-O core, surrounded by two nuclear burning layers: the inner one consisting of He, and the outer one of H. Those layers are in turn wrapped by a convective mantle and, further, by an atmosphere consisting of atomic and molecular gas, which is surrounded by a circumstellar envelope of gas and dust. 

The third dredge-up is the mechanism responsible for turning the abundance of AGB stars from O-rich into C-enriched \citep{ibenrenzini83}. Carbon-rich AGB stars are one of the most influential contributors to the enrichment of the interstellar medium, with dust made of amorphous carbon (amC) and silicon carbide (SiC). In their atmospheres carbon-bearing molecules, such as C$_2$, C$_3$, C$_2$H$_2$, CN, HCN, can be found. 

The evolution of the stars on the AGB is characterized by cooling, expansion and growing in brightness, burning the nuclear fuel faster and faster, and the star eventually begins to pulsate. The pulsation generates shock waves running through star's atmosphere, creating conditions of pressure and temperature suitable for dust formation. The sequence of pulsation and dust formation may drive a wind off the surface of the star into the interstellar space: when the opacity of amorphous carbon dust is high enough, the radiative pressure provides enough momentum to the grains to accelerate them and to drag along the gas by collisions, causing an outflow from the star (e.g. \citealp{Fleischer92}, \citealp{HoefnerDorfi}).

\cite{hofner03} describes this scenario with the solution of the coupled equations of hydrodynamics, together with frequency-dependent radiative transfer, including as well the time-dependent formation, growth, and evaporation of dust grains. The dynamic model atmospheres (DARWIN models, \citealp{hofner16}) that come from this code have successfully reproduced observations, e.g. line profile variations \citep{nowotny10} and time-dependent spectroscopic data \citep{Gautschy-Loidl04, walter13} of carbon-rich stars.

In our previous work \citep{rau15} we studied the atmosphere of the C-rich AGB star RU~Vir, comparing in a systematic way spectroscopic-, photometric- and interferometric-data with the grid of DARWIN models from \cite{mattsson10} and \cite{Erik14} .

The investigation of AGB stars using a combination of different techniques has been increasing over the last few years, while to date only a few interferometric observations of carbon stars have been directly compared with model atmospheres \citep{ohnaka07,paladini11,cruzal13,txpsc,vanbelle13}. From those, only very few have made use of time-dependent self-consistent dynamic models \citep{sacuto11,rau15}.		 
As suggested by \cite{hofner03}, this is the only way to acquire knowledge about the influence of the dynamic processes on the atmospheric structure, at different spatial scales. 

The purpose of this paper, is to extend our previous study on RU~Vir and to investigate the dynamic processes happening in the atmospheres of a set of C-rich AGB stars. To pursue this goal, we will compare predictions of DARWIN models, with observations by meaning of photometry, spectroscopy, interferometry. 
Long-baseline optical interferometry is an essential tool to study the stratification of the atmosphere, allowing to scan the regions of molecules and dust formation. 


The targets object of this study are the C-rich AGB stars: R~Lep, R~Vol, Y~Pav, AQ~Sgr, U~Hya, X~TrA, whose observations and parameters are described in Sect.~\ref{data}. \\
Section~\ref{geom} explores the geometry of the targets. Sect.~\ref{DARWIN models} introduces the self-consistent dynamic model atmospheres used, and presents their comparison with the different types of observables.  In Sect.~\ref{results} we present our results. Sect.~\ref{discussion} is a discussion of our results, including a comparison with the evolutionary tracks, and we conclude in Sect.~\ref{conclu} with perspectives for future work.


\section{Observational data}\label{data}
\subsection{The sample of targets}\label{sample}


Our sample consists of stars observed with the Very Large Telescope Interferometer (VLTI) of ESO’s Paranal Observatory with the mid-infrared interferometric recombiner (MIDI, \citealp{midi2003}) instrument, showing (1) an SiC feature in the visibility spectrum and (2) no evidence of asymmetry from differential phase (Paladini et al., in press). The stars can be grouped into Mira variables (\object{R~Lep}, \object{R~Vol}), Semiregular (\object{Y~Pav}, \object{AQ~Sgr}, \object{U~Hya}) and Irregular (\object{X~TrA}) stars \citep{GCVS}.

The main parameters of the stars, namely variability class, period, amplitude of variability, distance, and mass-loss rates, are shown in Table~\ref{table_starsparam}. For two stars, namely R~Vol and U~Hya, we are presenting new VLTI/MIDI data observed within the programs $090$.D-$0410$(A), $086$.D-$0899$(K). For the remaining stars, our data come from archive observations (Paladini et al., in press).

\begin{table*}[!htbp]
\caption{\label{table_starsparam} Main parameters of our target sample, adopted from the literature. 
\ldots indicates that no literature value is given.}
\centering
\begin{tabular}{lllllllll}
\hline
\hline
 Name  & Variability  &   $P$~\textsuperscript{a}   & $d$~\textsuperscript{b}  & $L_\text{bol}$~\textsuperscript{c}    &  $\dot{M}$~\textsuperscript{d}      &  $\dot{M}$~\textsuperscript{e}             & $\dot{M}$             & $\Delta{V}$~\textsuperscript{a}\\
 &  Type~\textsuperscript{a} &   [d] & [pc] & [L$_{\odot}$]  & $10^{-6}$[M$_{\odot}$/yr]  &  $10^{-6}$[M$_{\odot}$/yr] &     $10^{-6}$[M$_{\odot}$/yr] &              \\
\hline
R~Lep   &   M & 427 & $470^{+301}_{-122}$ & $8514$  &  $2.0 \pm 0.68$ & $0.70 \pm 0.35$  &  $0.93 \pm 0.19$~\textsuperscript{f}  & 6.2\\
R~Vol   &   M  & 454 & $880^{+149}_{-176}$& $8252$ & $2.9 \pm 0.68$  & $1.80 \pm 0.90$  & $1.99 \pm 0.34$~\textsuperscript{f} & 5.2\\
Y~Pav   &  SRb & 233 & $400^{+125}_{-77}$ & $5076$ & $2.8 \pm 0.96$  & $0.16 \pm 0.08$  & $0.23 \pm \ldots$~\textsuperscript{g}  &  1.7 \\
AQ~Sgr &  SRb & 200 & $330^{+95}_{-60}$  & $2490$ & $2.5 \pm \cdots $ & $0.25 \pm 0.12$ & $0.77 \pm \ldots$~\textsuperscript{g}  & 2.3 \\ 
U~Hya   & SRb & 450 & $208^{+35}_{-41}$ & $3476$  &  $0.5 \pm 0.05$  & $0.14 \pm 0.07$ &  $0.21 \pm \ldots$~\textsuperscript{g}  & 2.4 \\
X~TrA   &  Lb &  385 & $360^{+67}_{-49}$   & $8599$ & $0.5 \pm 1.05$  & $0.13 \pm 0.06$  & $0.18 \pm \ldots$~\textsuperscript{g}  & 1.4 \\
\hline
\hline
\end{tabular}\\
\textbf{Notes}. (a): \cite{GCVS}.. (b): the distances measurements come from \cite{vanleeuwen07}, except from R~Vol and U~Hya, which distances come from \cite{whitelock06}. (c): $L_\text{bol}$ is the bolometric luminosity, derived from the SED fitting. (d): \cite{loup93}. (e) \cite{schoier01}.  (f) \cite{whitelock06}. (g): \cite{bergeat05}.
\end{table*}

\subsection{Photometry}\label{phot}


We collected light curves for the $V$-band \citep{asas,aavso}, and the bands $J$, $H$, $K$, $L$ \citep{whitelock06, dirbe, lebertre92}. A mean value was derived for each filter with amplitudes derived from the variability (see Table~\ref{table_photometry} for details).
For the filters where no light curves are available (mainly $B$, $R$, and $I$), we averaged values collected from the literature. The errors were calculated as the standard deviation from those values. For the filters with only one value, without any literature associated error, an error of $20$\% was assumed.



 \subsection{Interferometry: MIDI data}\label{mididata}
All the targets of this study have been observed with the Auxiliary Telescopes (ATs) at VLTI. The observations were carried out with MIDI, which provides wavelength-dependent visibilities, photometry, and differential phases in the $N$-band ( $\lambda_{\text{range}} = [8,13]$ $\mu$m). 

Details on the data of R~Lep, Y~Pav, AQ~Sgr, X~Tra are given in Paladini et al. (in press), where the reader will also find the $uv$-coverage and the journal of the observations. 
 
The journal of observations of R~Vol and U~Hya is available in Appendix as online material (Table~\ref{tab_midi_observ_r_vol}, \ref{tab_midi_observ_u_hya}), together with the $uv$-coverages (Fig.~\ref{uv-coverage}). The calibrators used are listed below the corresponding science observation. 
The selection criteria for calibrators stars described in \cite{klotz12} were applied. The list of calibrators and their main characteristics are in Table~\ref{table_calibr}. 

The data reduction was made with software package MIA+EWS (V$2.0$ \citealp{jaffe04,ratzka07,midi2003}).  The size of the error bars is based on the calculated error in the visibilities. A conservative error of $10$\% on the visibilities is assumed in the case of a calculated error <$10$\%. The wavelength-dependent visibilities, shown in Figs.~\ref{interf-mira} and \ref{interf-semireg-irr}, exhibits the typical shape of carbon stars with dust shells containing SiC grains, which manifests its presence in the visibility minimum around $\sim11.3~\mu$m. The typical drop in the visibility shape between $8-9~\mu$m is caused by C$_2$~H$_2$~and~HCN molecular opacities.


\begin{table}[!htbp]
\caption{Calibrator list.}
\label{table_calibr}
\centering
\begin{tabular}{lllllll}
\hline
\hline
HD & Spectral type$^a$ & $F_\text{12}^\text{a}$ & $\theta^\text{b}$ \\
   &               & [Jy]        & [mas]        \\
\hline
32887           &       K4III           &        56.82  &       $5.90\pm0.06$     \\
81797		      &	K3II-III	          &	157.6	   &	    $9.14\pm0.04$	   \\
82668           &       K4/5III           &      73.10  &       $6.95\pm0.05$     \\
\hline
\hline
\end{tabular}
\tablefoot{(a)IRAS Point Source Catalog: \url{http://simbad.u-strasbg.fr/simbad/}.\newline 
(b)\url{{www.eso.org/observing/dfo/quality/MIDI/qc/calibrators}}}
\end{table}

 \begin{figure*}[!htbpbp]
\begin{center}
\resizebox{\hsize}{!}{
\includegraphics[width=0.9\hsize, bb=217 10 500 686, clip=true, angle=90]{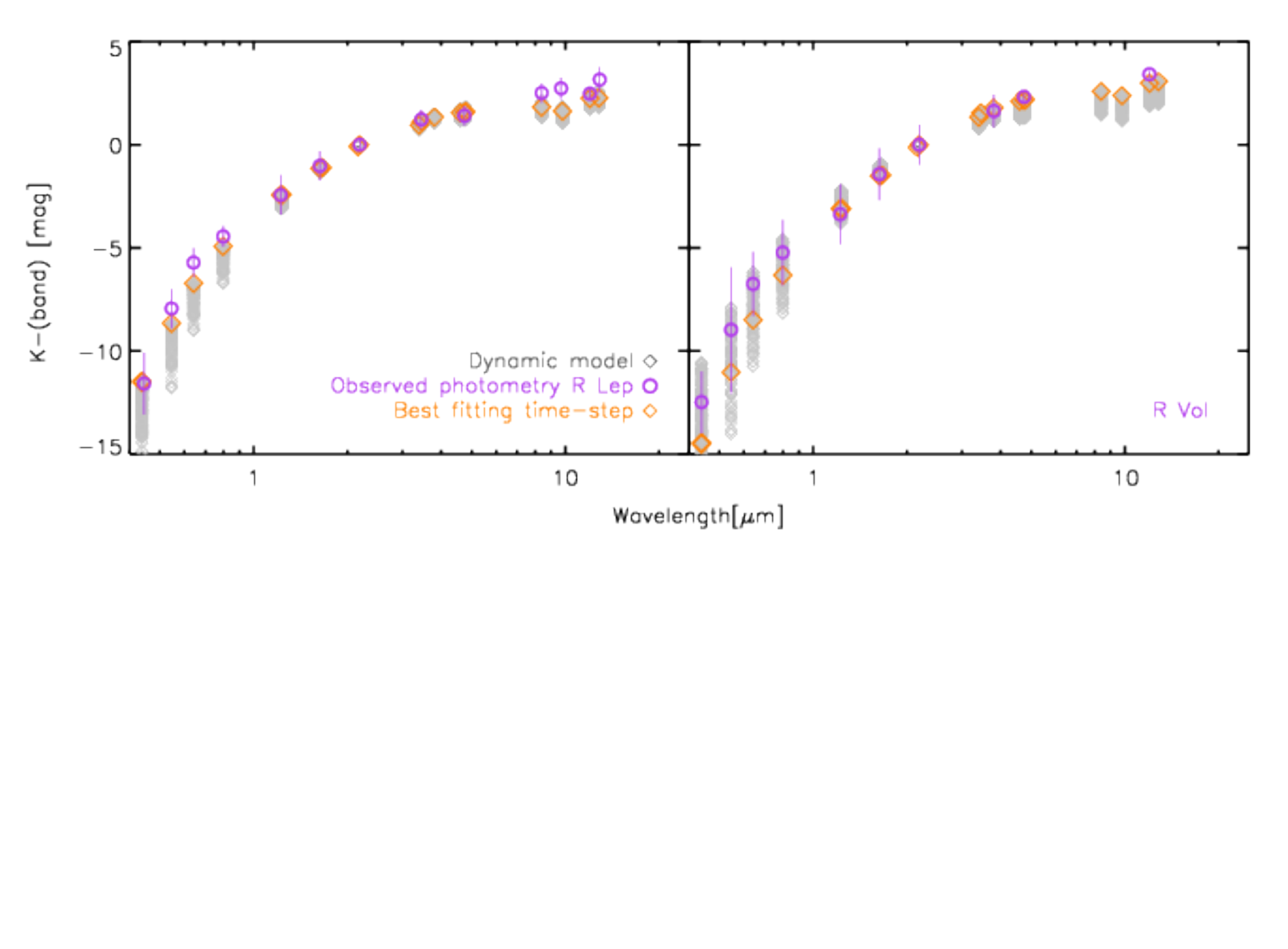}}
\caption{Photometric observations of \textbf{Mira} stars: R~Lep (left) and R~Vol (right). Observations (violet circles), compared to the DARWIN models synthetic photometry (grey diamonds). Orange diamonds show the best fitting time-steps of the two stars.}
\label{phot-mira}
\end{center}
\end{figure*}

   \begin{figure*}[!htbp]
\centering
\includegraphics[width=0.7\hsize, bb=0 0 504 684, clip=true, angle=90]{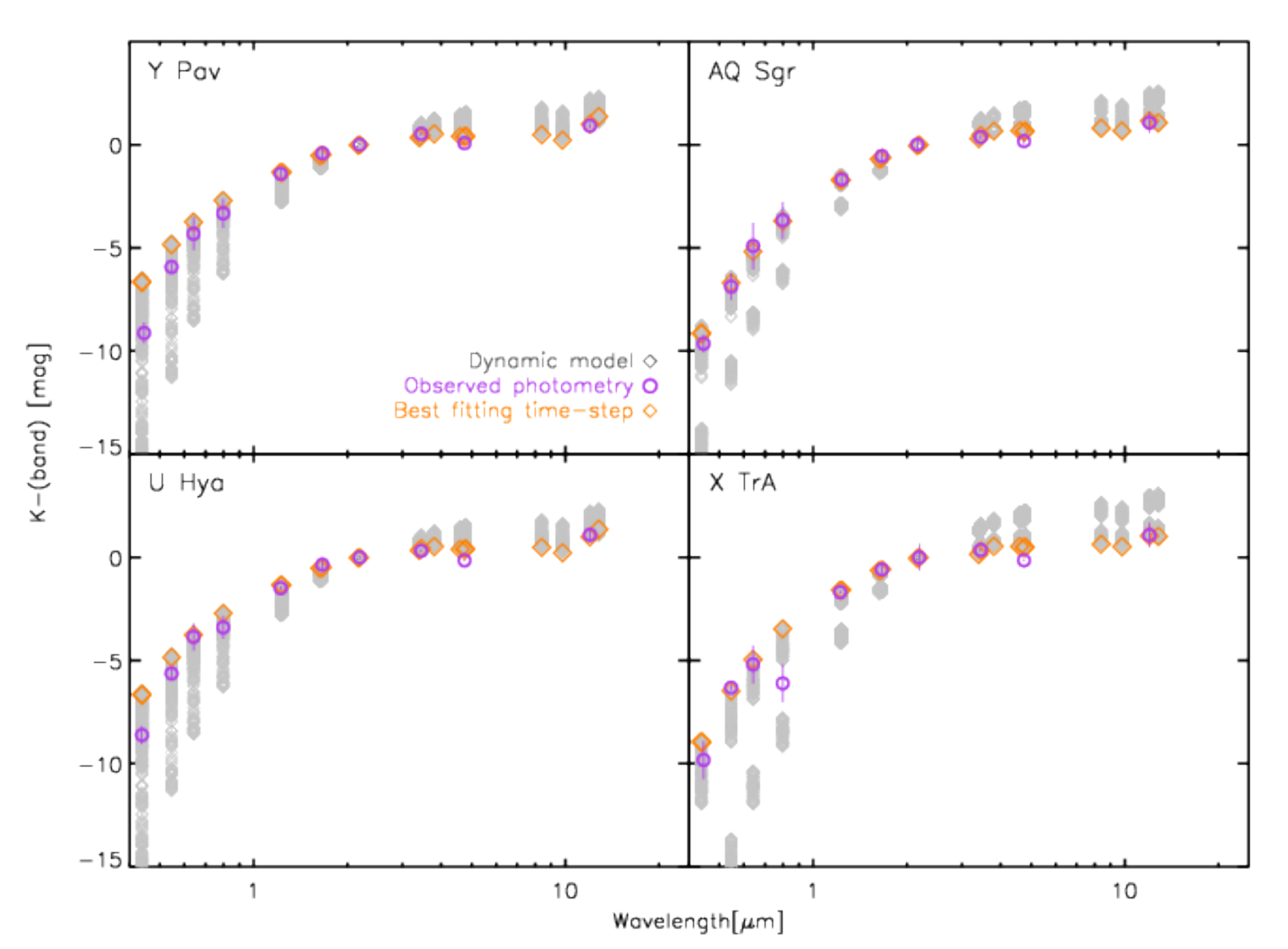}
\caption{Photometric observations of \textbf{SRb and Lb} stars: Y~Pav (upper left) and AQ~Sgr (upper right), U~Hya (lower left) and X~TrA (lower right). Observations (violet circles), compared to the DARWIN models synthetic photometry (grey diamonds). Orange diamonds show the best fitting time-steps of the four stars.}
\label{phot-semi-irr}
\end{figure*}

\section{Geometry of the environment}\label{geom}
As a first step, the MIDI interferometric data are interpreted with geometric models.   
To this purpose we used the GEM-FIND tool (GEometrical Model Fitting for INterferometric Data of \citealp{gemfind}) to fit geometrical models to wavelength-dependent visibilities in the $N$-band, allowing the constraint of the morphology and brightness distribution of an object. The detailed description of the fitting strategy and of the $\chi^2$ minimization procedure can be found in \cite{klotz12}. 

U~Hya has only one visibility spectrum available (see $uv$-coverage in Fig.~\ref{uv-coverage}, right side) therefore only Uniform Disc (UD) and Gaussian models can be applied. By fitting the data we derived a UD-equivalent diameter at $8$ and $12~\mu$m, of respectivelly $\theta_{8} = 23.89 \pm 2.54~$mas and $\theta_{12} = 39.26 \pm 2.64~$mas and Gaussian full width half maximum (FWHM)  at $8$ and $12~\mu$m of $14.60 \pm 1.68~$mas and $24.48 \pm 1.87~$mas respectively.



Two MIDI data points are available for R~Vol ($uv$-coverage shown in Fig.~\ref{uv-coverage}, left side). The angular diameters derived from the fit are: $\theta_{8} = 26.38 \pm 0.17~$mas and $\theta_{12} = 33.45 \pm 0.36~$mas for a circular UD fit, and $\theta_{8} = 17.88 \pm 0.36~$mas and $\theta_{12} = 24.48 \pm 0.60~$mas for a fit with a circular Gaussian model.

 
Geometric modelling for the other stars of our sample are presented in Paladini et al. (in press) For the discussion and interpretation we refer the reader to the values published in their Table~$4$.

\section{Dynamic Models Atmospheres}\label{DARWIN models}

\subsection{Overview on the DARWIN models}

Our observational data are compared with synthetic observables obtained from the grid of DARWIN models presented in \citet{mattsson10} and \citet{Erik14}, and model spectra. A detailed description of the modelling approach can be found in \citet{HoefnerDorfi}, \citet{hofner99},  \citet{hofner03}, \citet{hofner16}. Applications to observations are described in \citet{loidl99}, \citet{Gautschy-Loidl04}, \citet{nowotny10}, \citet{walter2}, \citet{sacuto11}, \citet{rau15}.

Those models result from solving the system of equations for hydrodynamics, and spherically symmetric frequency-dependent radiative transfer, plus equations describing the time-dependent dust formation, growth, and evaporation. 
The initial structure of the dynamic model is hydrostatic. A ``piston'' simulates the stellar pulsation, i.e. a variable inner boundary below the stellar photosphere. The "method of moments'' \citep{gauger90,gail88} calculates the dust formation of amorphous carbon. 
 
The main parameters characterising the DARWIN models are: effective temperature $T_{\text{eff}}$, luminosity $L$, mass $M$, carbon-to-oxygen ratio $C/O$, piston velocity amplitude $\Delta_{\text{u}}$, and the parameter $f_\text{L}$ used in the calculations to adjust the luminosity amplitude of the model. The emerging proprieties of the hydrodynamic calculations are the mean degree of condensation, wind velocity, and the mass-loss rate. A set of ``time-steps'' describe each model, corresponding to the different phases of the stellar pulsation. 

The synthetic photometry, synthetic spectra and synthetic visibilities are computed using the COMA code and the subsequent radiative transfer \citep{aringer00, aringer09}. The synthetic photometry is derived integrating the synthetic spectra over the selected filters mentioned in Sect.~\ref{phot}. Starting from the radial temperature-density structure at a certain time-step taken from the dynamical calculation, and considering the equilibrium for ionization and molecule formation, all the abundances of the relevant atomic, molecular, and dust species were calculated. The continuous gas opacity and the strengths of atomic and molecular spectral lines are subsequently determined assuming local thermal equilibrium (LTE). The corresponding data, listed in \cite{cristallo07} and \cite{aringer09} are consistent with the data used for constructing the models. 

The amount of carbon condensed into amorphous carbon (amC), in g/cm$^3$, as a direct output of the calculations, is taken from the models. amC dust opacity is treated consistently (\citealp{roleau_and_martin} in small particle limit(SPL)), and further details on the dust treatment are given in \cite{Erik14}. SiC is added, artificially, a posteriori with COMA.~
Following \cite{rau15} and \cite{sacuto11}, the percentage of condensed material is distributed in this way: $90$~\% amorphous carbon, using data from \cite{roleau_and_martin}, and $10$\% silicon carbide, based on \cite{pegourie}. Some experiments that change this configurations are presented in Sect.~\ref{discussion}.

All grain opacities are calculated for the SPL, in order to be consistent\footnote{An inconsistent treatment of grain opacities causes larger errors in the results than does using the small particle limits approximation.} with the model spectra from \cite{Erik14}. The assumed temperature of the SiC particles equals the one of amC; this is justified, since the overall distribution of the absorption is quite similar for both species, except for the SiC feature around $11.3~\mu$m. As a consequence, the addition of SiC would also not cause significant changes in the thermal structure of the models. Since the SPL is adopted, the effects of scattering are not included, as they are neglegible in the infrared.

\subsection{The fitting procedure}\label{fitting_procedure}

Generating one synthetic visibility profile for each of the approximately $140$~$000$ time-steps of the DARWIN models grid, and for each baseline configuration of our observations, would be extremely time-consuming from a computational point of view. Therefore a simultaneous fitting of the three types of observables was excluded a priori, instead implementing the procedure described as follows. 





\textbf{First}, the photometric observations were compared to the synthetic DARWIN models photometry. In the case of R~Lep, also the spectro-photometric data were fitted. 
A $\chi^2$ minimization was performed over the available literature photometric data, for each of the $540$ models of the grid, with a total of approximately $140~000$~time steps. The best fitting photometry model, with a corresponding best fitting-photometry time-step is listed in Table~\ref{tab_param_dyn}.


We would like to note that it is beyond the intent of this paper to model individual phases in terms of photometry and of interferometry. Indeed, the fit of the averaged observed photometry to the single time-steps of the models was done
only to the aim of pre-selecting a model for the subsequent interferometric comparison. 

As mentioned above, it is not feasible to calculate the synthetic intensity profiles and visibilities for each of the grid's ~140000 time-steps. 
Furthermore, data longwards of $2~\mu$m are usually single epoch (i.e. one observing date) and the data for the shorter wavelengths are often only some few measurements and usually from several light cycles. 
So the observations correspond to a random mixture of phases, both with respect to wavelength (phase coverage of different filters quite different) and with respect to cycles. Thus in our case the mean of the available photometry will be different from the mean SED (which corresponds to a mean pulsational phase). 

\textbf{As next step}, we produced the synthetic visibilities, following the approach of \cite{davis00}, \cite{tangoanddavis02} and \cite{paladini09}. They are calculated as the Hankel transformation of the intensity distribution $I$, which results from the radiative transfer. We then compared them to the interferometric MIDI data of each star. 
For the time-computational reasons mentioned before, we produced the synthetic visibilities only for the best fitting photometric model, i.e. for all the time-steps belonging to that model. 

Concerning interferometry, no variability in the $N$-band was observed for the semiregular star R~Scl \citep{sacuto11} and for the irregular star TX~Psc \citep{txpsc}. Paladini et al. (in press),  where part of our targets are studied (R~Lep, Y~Pav, AQ~Sgr, X~TrA), concluded that interferometric variability is of the order of $10$~\% or even less. Following this results, and considering that $10$~\% is the typical error of our interferometric dataset, we assume that no interferometric variability is present and we combine the observations as representative of one single snapshot of the star. Indeed, if data for more than one epoch are available, then all data will be combined for the fit. Generally we have MIDI data for only one or two epochs, with typically a small or no overlap in baseline and position angle as would be necessary in order to check for interferometric variability. Therefore fitting individual data for single epochs with single time steps would have notably reduced the significance of the fits. The sparse coverage in variability phase and $uv$-space of the MIDI observations also did not indicate a fit of averaged observations with averaged model visibilities. 
The interferometric $\chi^2$ values of the best fitting timesteps (Table~\ref{tab_param_dyn}) are provided for completeness and to guide the discussion. For readability of the figures involving model visibilities, only the best time-step is shown. The assumption of small interferometric variability and the range of model visibilities are discussed in Sect.~\ref{interfer-variability}.

In the following paragraphs we present the results of the comparison of the DARWIN models with the spectro-photometric and interferometric data for each single star. One example of the confrontation of the intensity profile and visibility vs. baseline at two different wavelengths, namely $8.5~\mu$m for the molecular contribution and $11.3~\mu$m for the SiC feature, is shown in Fig.~\ref{interf-rlep-visbase} (see Appendix for the other stars).




 \begin{figure*}[!htbp]
   \centering   
   \includegraphics[width=0.7\hsize, bb=0 0 504 684, clip=true, angle=90]{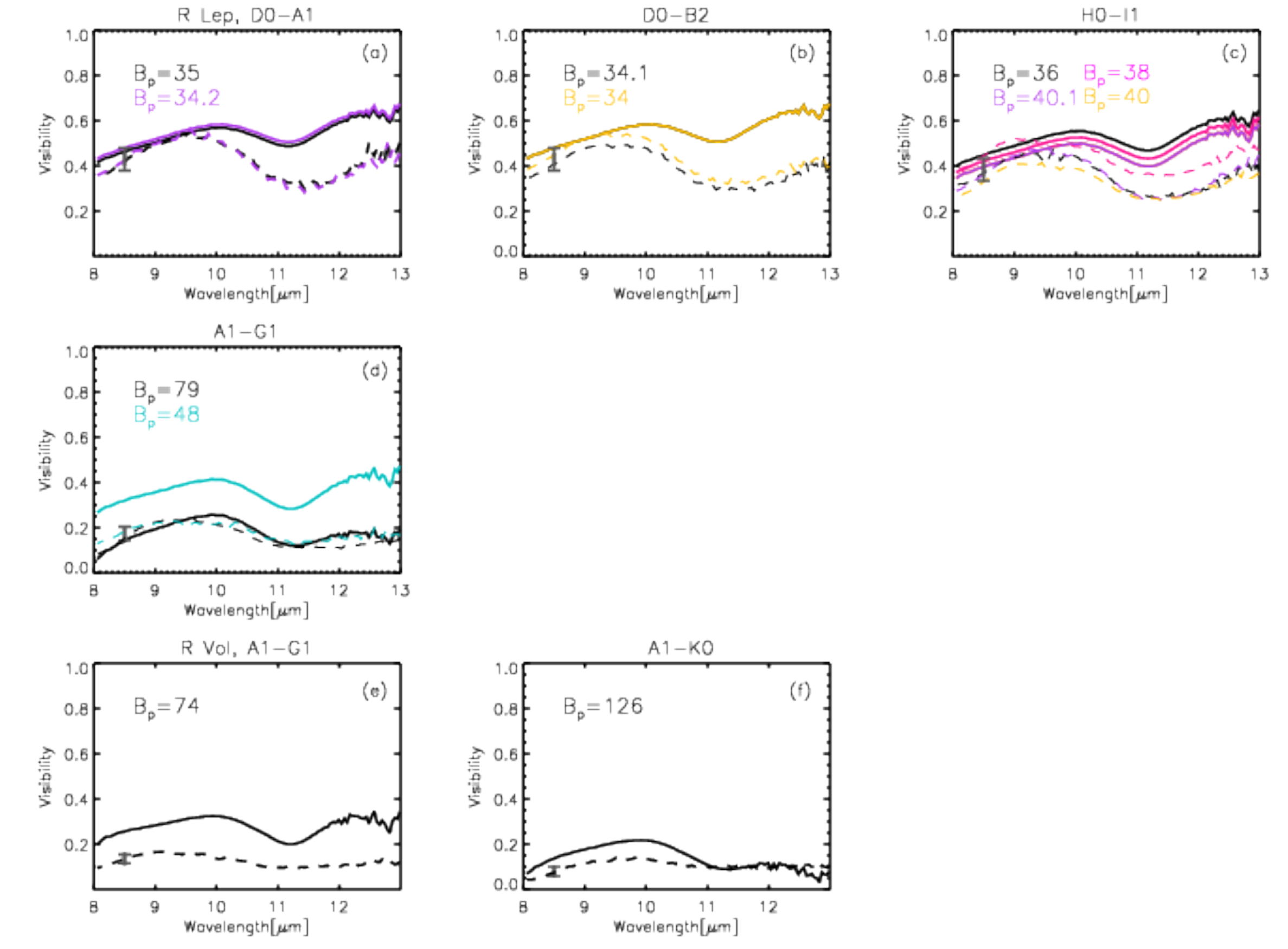}
    \caption{\label{interf-mira} Visibility dispersed over wavelengths of the Mira variables of our sample. Models are plotted in full line, observations in dashed lines, at the different projected baselines (see color legend). The stars are identified in the title. The six panels show R~Lep dispersed visibilities at the baseline configuration D0-A1 (a), D0-B2 (b), H0-I1 (c) and A1-G1 (d), as also marked in the plot titles. R~Vol dispersed visibilities are at the baseline configuration A1-G1 (e) and A1-K0 (f). Errorbars are of the order of $10$~\%, and a typical error-size bar is shown in grey in each panel, overlapping with the data at $8.5~\mu$m. In panel (a) the two models full lines are overlapping, as the two observations lines. In panel (d) the two lines of the observations are overlapping, and the model at B$_p = 79$ also lies on top of them.} 
\end{figure*}

\begin{figure*}[!htbp]
   \centering   
   \includegraphics[width=0.7\hsize, bb=0 0 504 684, clip=true, angle=90]{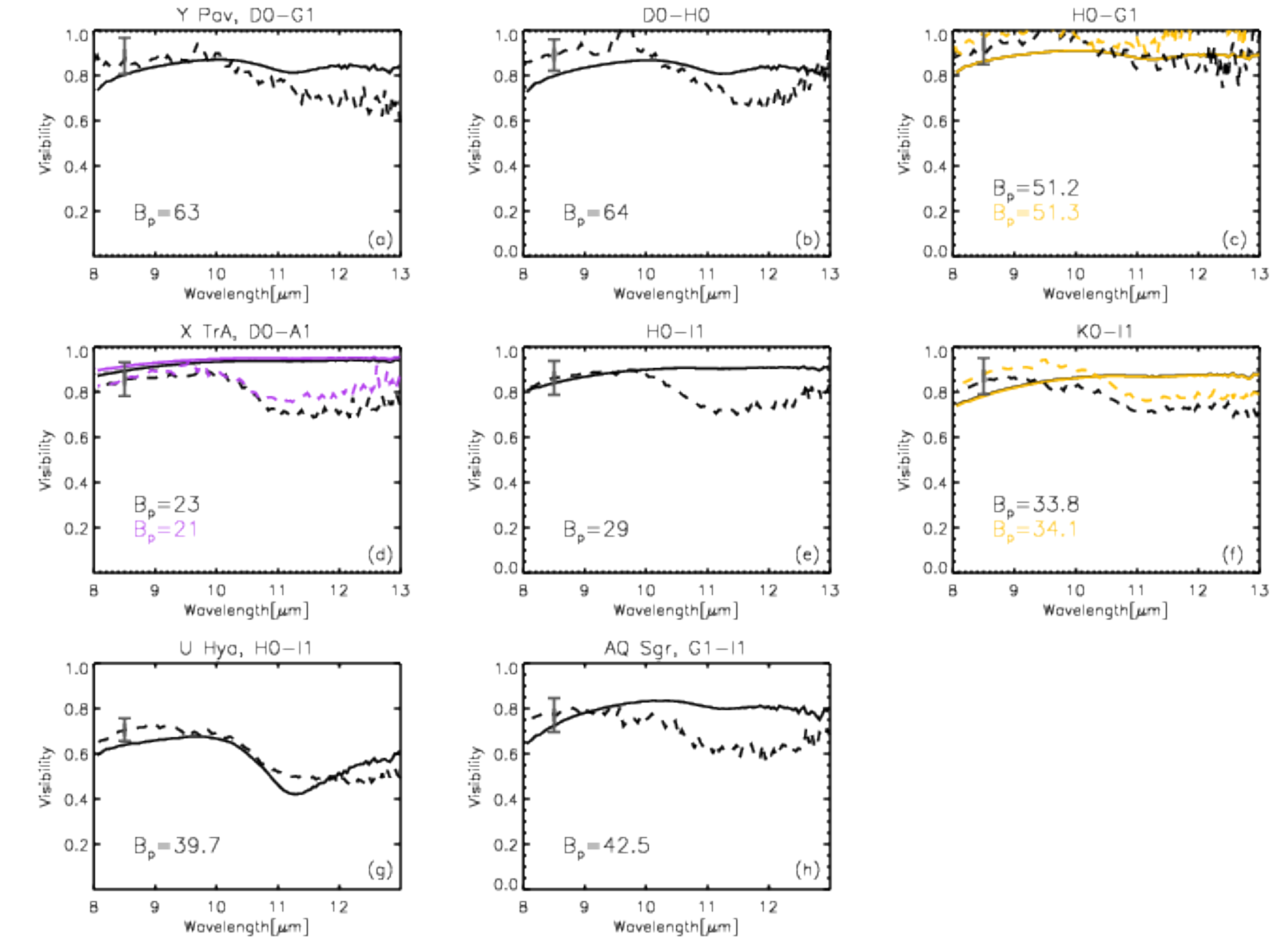}
\caption{\label{interf-semireg-irr} Same as Fig.~\ref{interf-mira}, for the Semiregular and Irregular stars of our sample: Y~Pav in panels (a), (b), (c); X~TrA in panels (d), (e), (f); U~Hya in panel (g) and AQ~Sgr in panel (f). Errorbars are of the order of $10$~\%, and a typical error-size bar is shown in grey in each panel, overlapping with the data at $8.5~\mu$m. Panels (c), (d), (f) they show the two full lines (models) that overlap.} 
\end{figure*}

\section{Results}\label{results}

The DARWIN models fits with our three different types of observations, lead to results which are described in this section, for Mira, Semiregular and Irregular stars. Please refer to Sect.~\ref{discussion} for a detailed discussion on our results.

The results of the fit, namely the $\chi^2$, are shown in Table~\ref{tab_param_dyn}. The main parameters that characterize the models, as described in Sect.~\ref{DARWIN models} are listed together with the resulting properties of the DARWIN models, such as the mean mass-loss rate $\dot{M}$. 
No assignment of MIDI phases can be done for the Semi-regular and Irregular variables, due to the non regular nature of their light curves and also the sometimes poor phase coverage of the light curves. 
We want to remark that all out targets show no evidence of asymmetry from differential phase in the MIDI data (see also Paladini et al., in press).

The best fitting models of Y~Pav, AQ~Sgr, U~Hya and X~TrA resulted, at first, in models without mass-loss. Since those stars show presence of mass-loss in the literature (see Table~\ref{table_starsparam}), we decided to perform a selection a priori, choosing from the whole grid of $560$ models, only the ones allowing for wind formation, i.e. having a condensation factor $f_\text{c} > 0.2$. This results in a sub-grid of $168$ models, among which we performed our analysis for the Semiregular and Irregular stars. 
We will, however, discuss also the fits with the windless models for these stars in Sect.~\ref{semiandirregular}.

Based on our findings, some general statements can be made: overall, the $\chi^2$ from SED fitting of non-Miras is higher than the one obtained for Mira variables (Table~\ref{tab_param_dyn}). We also found that the Miras interferometric observations show the SiC feature shallower than the one produced by the DARWIN models. 

\subsection{Mira stars}\label{mira}

\begin{figure*}[!htbpbp]
\begin{center}
\resizebox{\hsize}{!}{
   \includegraphics[width=\hsize, bb=0 0 504 684, clip=true, angle=90]{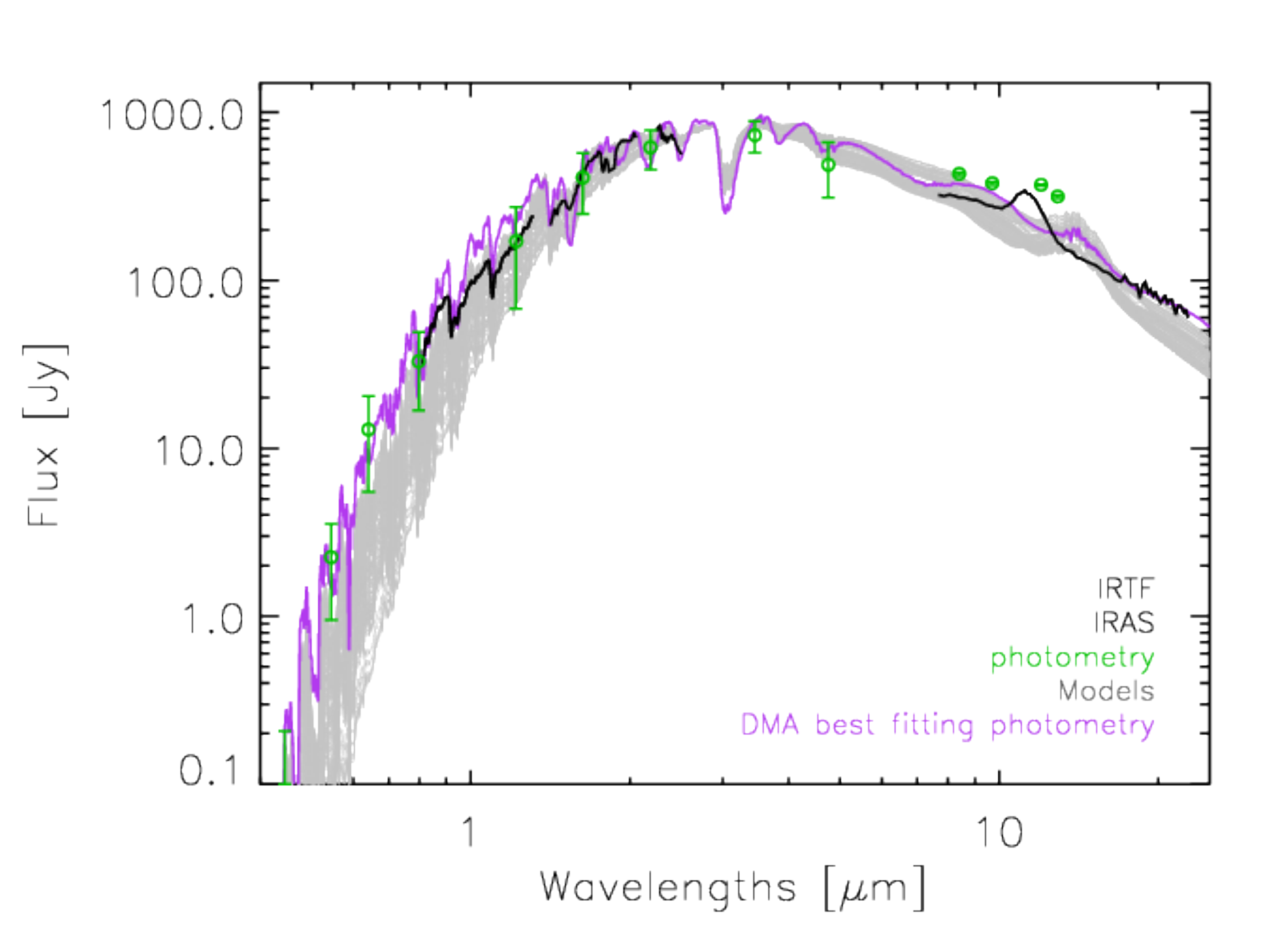}}
    \caption{\label{spec-rlep} Observational spectro-photometric data of \textbf{R~Lep}, compared with the synthetic spectrum of the best fitting time-step (violet). Photometry is plotted in green circles, while IRAS (\citealp{iras-satellite}) and NASA/IRTF \citep{irtf} spectra are plotted black lines, to the purpose to check qualitatively the photometric fit. The spectrum of the DARWIN models for which the synthetic photometry fits best the corresponding observational data is shown, in violet.}
\end{center}
\end{figure*}

The spectroscopic and photometric data of R~Lep agree well with the model predictions, as can be seen in Fig.~\ref{spec-rlep} (in which the IRAS spectrum has been over-plotted for qualitative comparison reasons) and Fig.~\ref{phot-mira}, left panel. The small differences at wavelength shorter than $1~\mu$m are discussed later in Sect.~\ref{diff-1mu}. The model SED shows an emission bump around $14~\mu$m, which is not seen in the observed spectrum, as noticed also by \cite{rau15}. The origin of this feature predicted by the model, is due to C$_2$H$_2$ and HCN, as mentioned by \cite{loidl_phdthesis}, and discussed in Sect~\ref{visib-slope}.

\begin{figure*}[!htbp]
\begin{center}
\resizebox{\hsize}{!}{
\includegraphics[width=\textwidth, bb=0 0 504 684, angle=90]{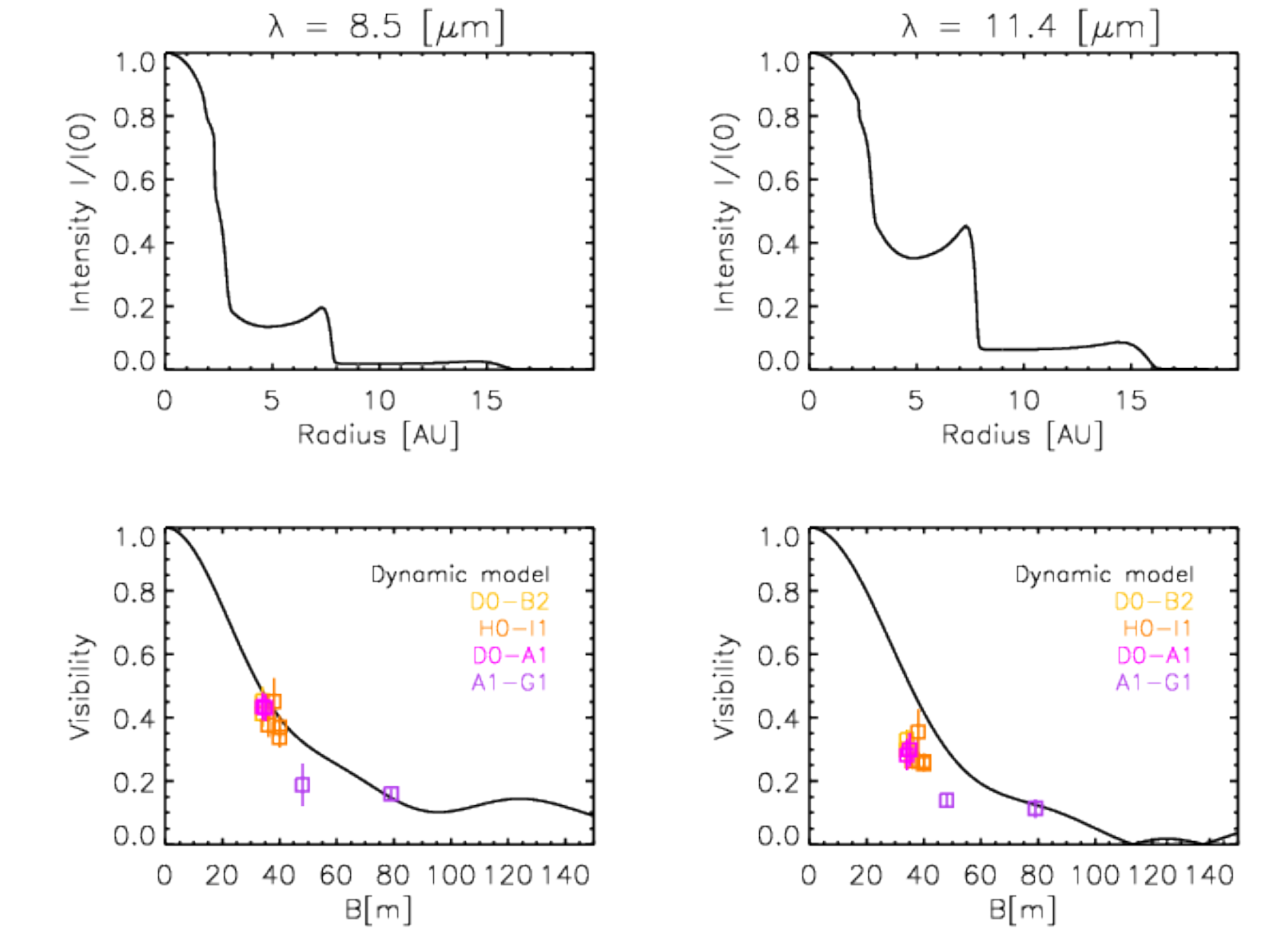}}
\caption{Interferometric observational MIDI data of R~Lep, compared with the synthetic visibilities based on the DARWIN models. \textbf{Up}: intensity profile at two different wavelengths: $8.5~\mu$m and $11.4~\mu$m. \textbf{Down}: visibility vs. baseline; the black line shows the dynamic model, the colored symbols illustrate the MIDI measurements at different baselines configurations.}
\label{interf-rlep-visbase}
\end{center}
\end{figure*}

The good fit is confirmed by a $\chi^2$ of $0.99$ and $1.01$ for photometry and interferometry respectively (Table \ref{tab_param_dyn}). 
R~Lep interferometric data 	are shown in Fig.~\ref{interf-rlep-visbase} and Fig.~\ref{interf-mira} (upper panels). In the latter, the typical SiC shape around $11.3~\mu$m is visible. This shape is reproduced by the models, and their difference in visibility level at wavelength longwards of $10~\mu$m will be discussed in Sect.~\ref{discussion}.

The R~Vol photometric data show good agreement at all wavelength ranges, well within the error bars (see Fig.~\ref{phot-mira}, right panel). The interferometric data of R~Vol are taken with long baselines ($B_\text{p} =74$~m and $B_\text{p} =126$~m), and cover visibility values between $0.05$ and $0.15$. Overall, the observations taken with the $126$~m baseline (Fig.~\ref{interf-mira}) are reproduced by the model in the wavelength range between $11$ and $13~\mu$m. However the model predicts higher visibilities for the  $74$~m baseline.

   

In summary, the Mira stars exhibit a visibility vs. wavelength profile always flatter than the models, and agrees better at wavelengths shorter than $10~\mu$m. A similar fixnding was reported by \cite{sacuto11} for R~Scl.

\begin{figure*}[!htbp]
\begin{center}
\resizebox{\hsize}{!}{
\includegraphics[width=\textwidth, bb=0 0 504 684, angle=90]{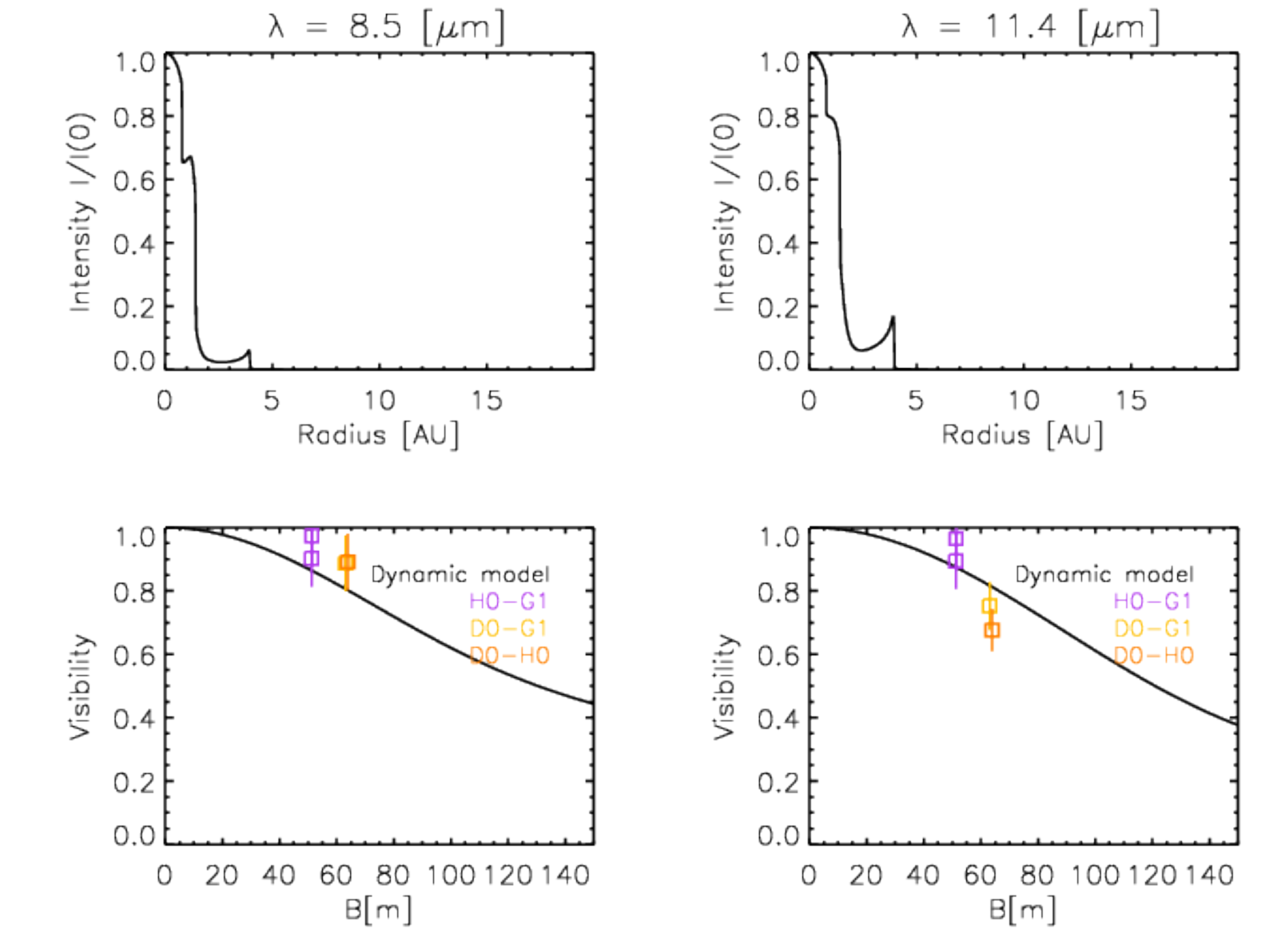}}
\caption{Same as Fig.~\ref{interf-rlep-visbase}, but for Y~Pav.}
\label{interf-ypav-visbase}
\end{center}
\end{figure*}

 \subsection{Semi-regular and Irregular Variables}\label{semiandirregular}

As already mentioned above, at first the best fitting models of Semi-regular and Irregular stars resulted in those without mass-loss. The corresponding classes resulted in pp (periodically pulsating) for Y~Pav, U~Hya and X~TrA, and pn (non-periodic) for AQ~Sgr. The parameters of those models without mass loss are indicated in Table~\ref{tab_param_dyn} for comparison. 
In general, the $\chi^2$ of the photometry of the Semi-regular and Irregular stars is higher than for the Miras (see Table~\ref{tab_param_dyn}). Compared to the windless models, the fit of SEDs with mass-loosing models ($f_\text{c}~>~0.2$) did not lead to a better fit  for wavelengths shorter than $1~\mu$m. Furthermore, the total visual amplitudes of these models, which are mostly due to variable dust extinction \citep{walter2} are markedly larger than the observed ones. 
On the other hand, all these models have either episodic or multi-periodic mass loss which leads to less regular or multi-periodic synthetic light curves, similar to the observed characteristics. Looking at individual cycles, the visual amplitudes are closer to the observed ones.
The visibility slope of the models with mass-loss agrees better than for Miras, and the visibility level is always high, except for U~Hya (for detailed plots please refer to Fig.~\ref{interf-ypav-visbase}, \ref{interf-uhya-visbase}, \ref{interf-aqsgr-visbase}, \ref{interf-xtra-visbase}). We want to underline that in comparison, the windless models are too compact and also lack the SiC signatures observed in the visibilities. This can be seen in Appendix, Fig~\ref{nomassloss-ypav} and Fig.~\ref{nomassloss-xtra}, that show visibilities vs. wavelength of the best fitting models without mass loss respectively for Y~Pav (semi-regular) and X~TrA (irregular). We thus decided,  based on the observed mass loss and the significantly better fit of the visibilities, to consider the mass-loosing models for the remaining analysis.


The synthetic SED from the DARWIN models of Y~Pav agrees with the observations well, except for the $B$ filter (see Fig.~\ref{phot-semi-irr}). The problem of having the $B$ filter photometric data off the fit, also appearing for some of the other targets, manifests itself also in the SED of Y~Pav, and a likely reason of this is discussed in Sect.~\ref{discussion}. The interferometric data show high visibility level at all three Y~Pav baseline configurations. The models agree in level with the MIDI observations, and their difference in shape is discussed in Sect.~\ref{discussion}.

The synthetic photometry of AQ~Sgr fits the data well within the error bars. The absolute visibility level of the MIDI data is in agreement with the models, but the SiC feature shape is not as pronounced in the models as in the observations.

The synthetic DARWIN models SED of U~Hya is in good agreement with the observations. The small discrepancy shorter than $1~\mu$m is discussed in Sect.~\ref{discussion}. The synthetic visibilities seems to reproduce well, within the error bars, the shape and level of the MIDI U~Hya observations.

Since at first the photometry of X~TrA in Fig.~\ref{phot-semi-irr} lower right panel, had the value in the filter $I$ particularly offset compared to the overall fit, we performed a new fit excluding those values. Since the ``new’’ best fitting model has no mass-loss, we repeated the fitting procedure again following the selection of models explained in Sect.~\ref{fitting_procedure}. The reduced-$\chi^2$ obtained for the SEDs following this procedure is $6.2$.   
The results are shown in Table~\ref{tab_param_dyn} and Fig.~\ref{phot-semi-irr}. There is a good agreement between models and MIDI observations, and the discrepancy in shape is examined in Sect.~\ref{discussion}.

\begin{figure}[!htbp]
\begin{center}
\resizebox{\hsize}{!}{
\includegraphics[width=\textwidth, angle=90,bb=251 353 490 668]
{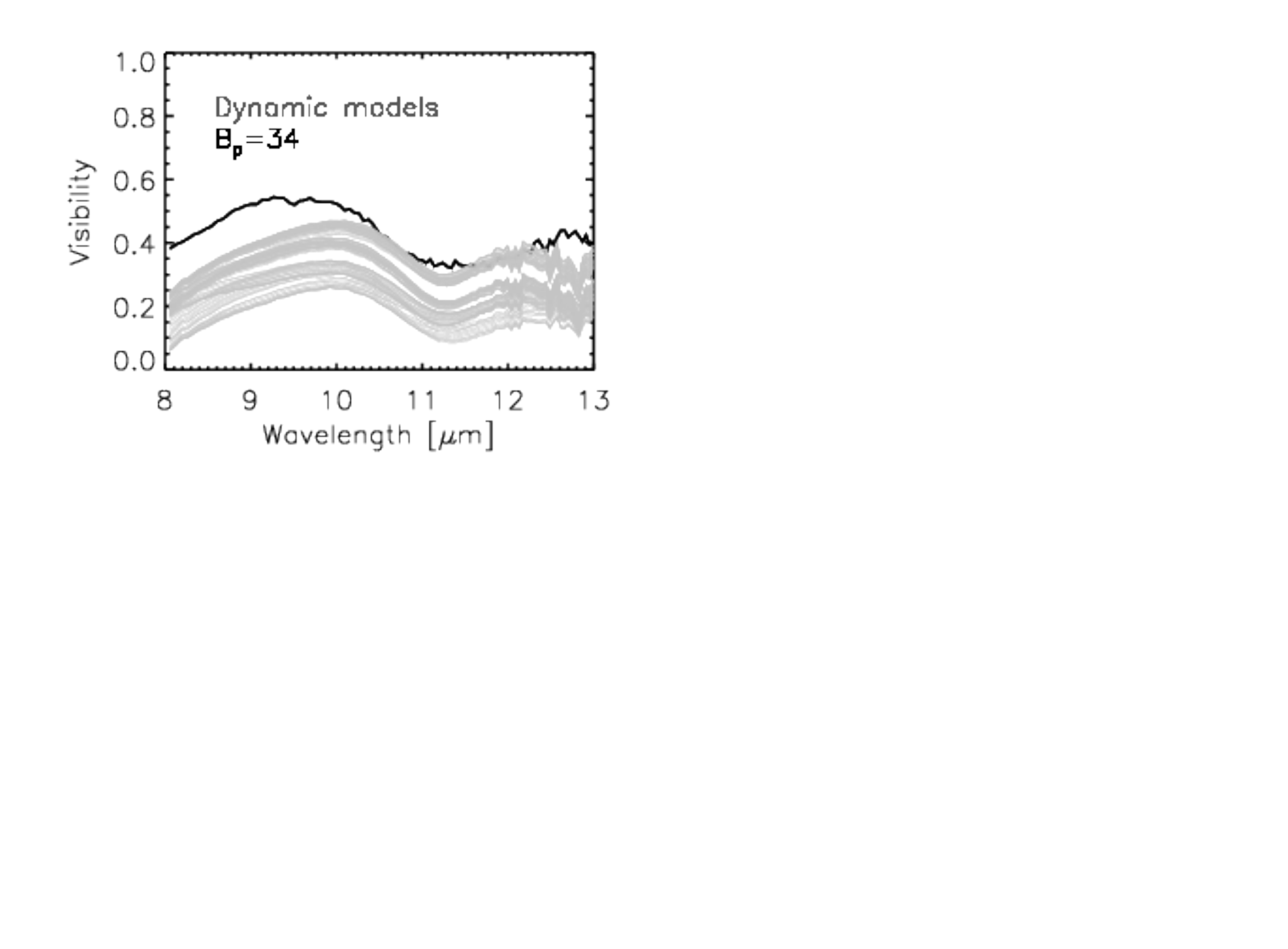}}
\caption{Visibilities dispersed in wavelengths for the shorter baseline (B$_p = 34$~m) of R~Lep observations, in black. The grey lines illustrate the range in visibility of the model's time-steps.}
\label{rlep_range_bp34}
\end{center}
\end{figure}

\begin{figure}[!htbp]
\begin{center}
\resizebox{\hsize}{!}{
\includegraphics[width=\textwidth, angle=90,bb=251 353 490 668]{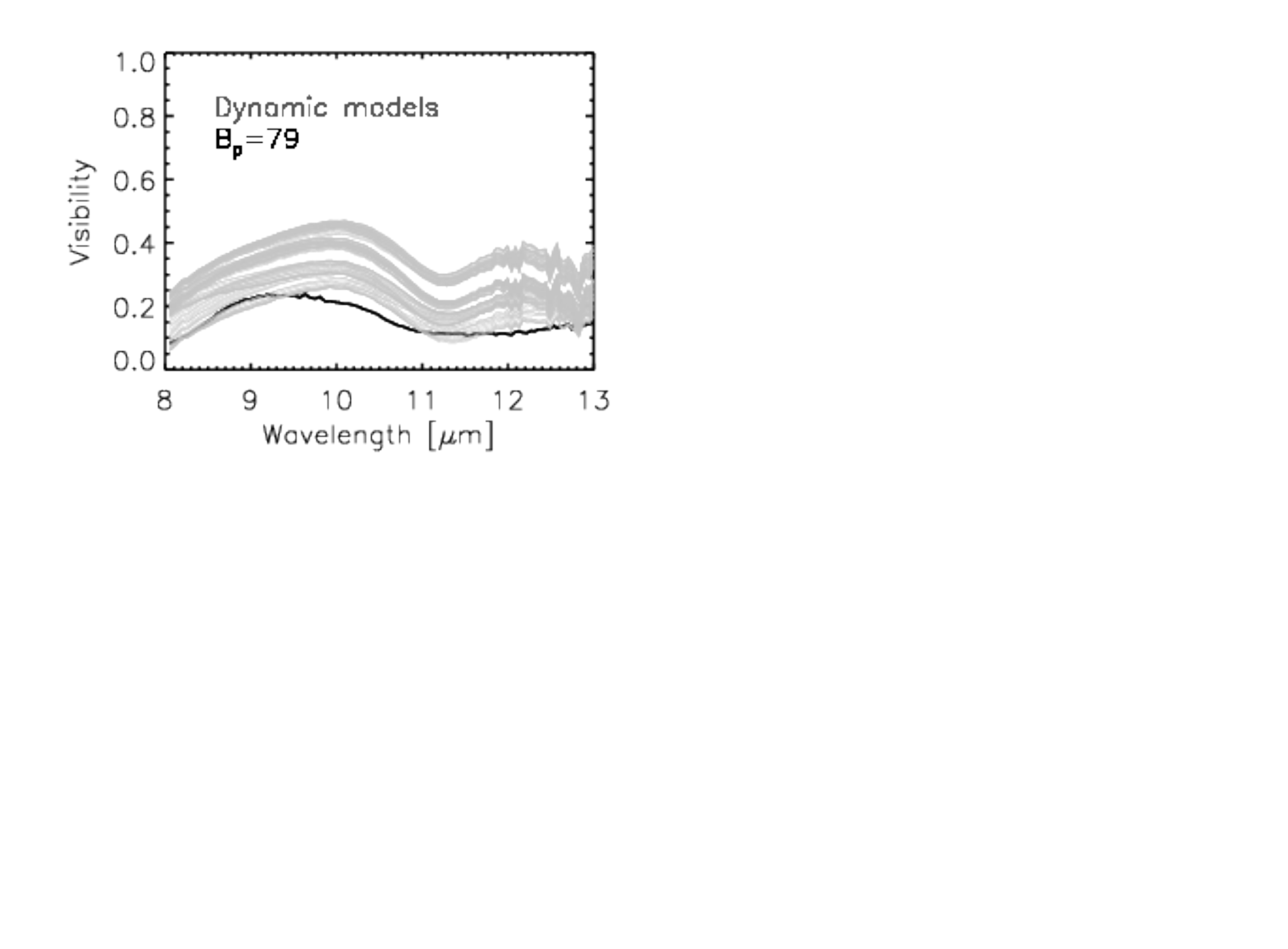}}
\caption{Same as Fig.~\ref{rlep_range_bp34}, but for the longest baseline of R~Lep observations (B$_p = 79$~m).}
\label{rlep_range_bp79}
\end{center}
\end{figure}

\begin{figure*}[!htbp]
\begin{center}
\resizebox{\hsize}{!}{
\includegraphics[width=0.7\textwidth, angle=90,bb=245 13 501 667]{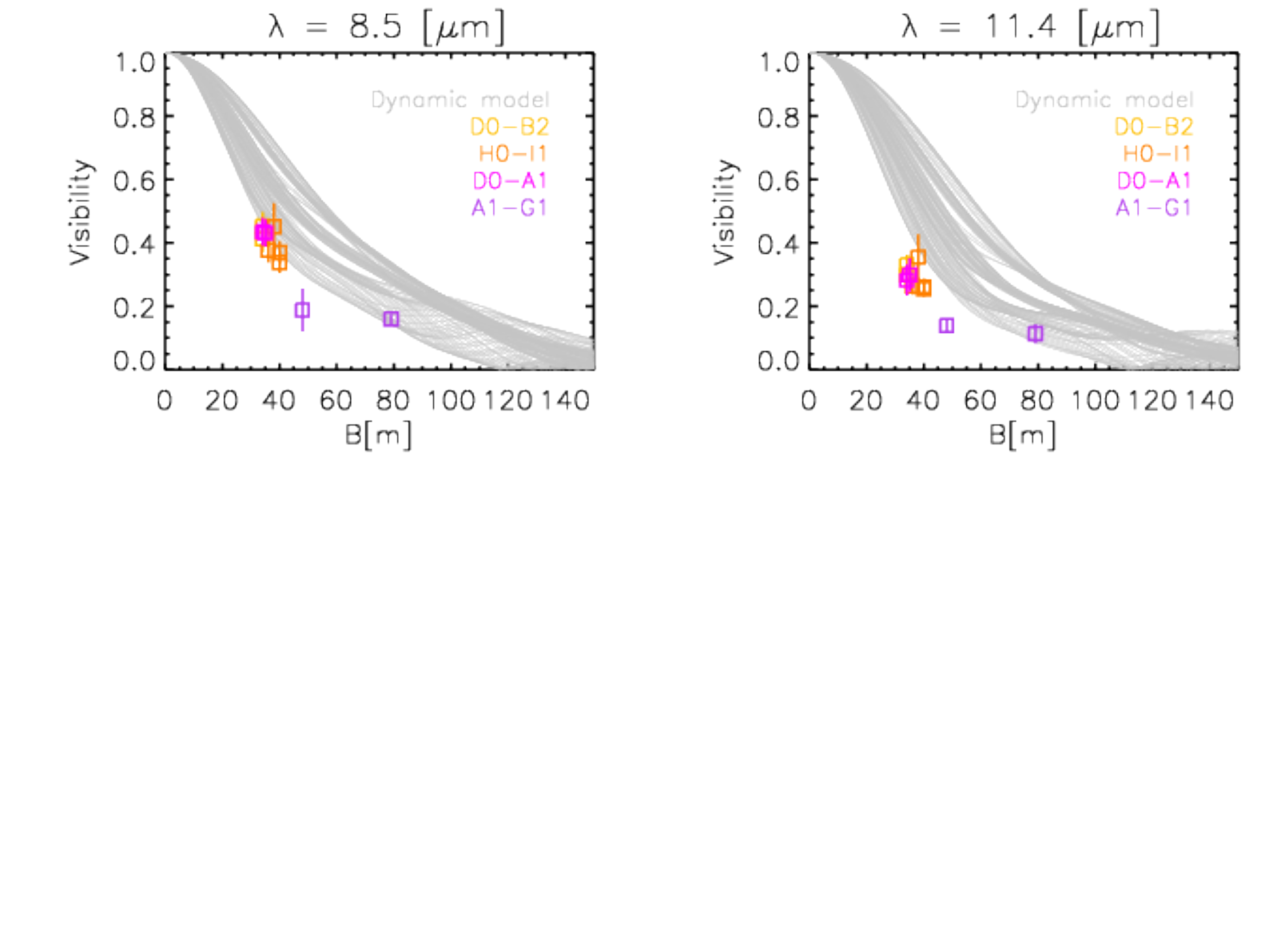}}
\caption{Visibilities vs. baselines of R~Lep model, illustrating the range of available synthetic time-steps (in grey), at two picked wavelength: $8.5~\mu$m and $11.4~\mu$m. The colored squares are the R~Lep observations.}
\label{rlep_range_visbase}
\end{center}
\end{figure*}

\begin{table*}
\centering
\caption{\label{tab_param_dyn} Summary of the best fitting model for each type of observation: photometry, spectroscopy and interferometry. Listed are the corresponding values of the $\chi^2$, and the parameters of the models.}
\begin{tabular}{lllllllllllll}
\hline
\hline
 & $T_{\text{eff}} $ & $\log~L_\text{bol}$ & $M$ & $P$ & $\log~g$ & $C/O$ & $\Delta u_{\text{p}}$ & $f_{\text{L}}$ & $\dot{M}$  & $\lambda_{\text{fit range}}$ & ${\chi^2}_{\text{red}}$  \\
 &$[K]$ & [L$_{\odot}$] & [M$_{\odot}$] & [d] &   &  &  &   & $10^{-6}$[M$_{\odot}$/yr] &   [$\mu$m]  &   & \\
\hline
\bf{R~Lep}  &              &                 &   &   &    &  &      &      &      &    &    \\
\hline
Spectr & 3000&	3.85&	1.0&  390&  -0.57& 1.69&	  6&  2&  2.45&  [0.805-5.06] &  0.99   \\
\hline
Photom & 2800&	3.85&	1.00& 390&    -0.69& 1.69&  6&  1&  2.24&   [0.4-25.0]  &    1.03   \\
\hline
Interf &   2800&	3.85&	1.00& 390&    -0.69& 1.69&  6&  1&  2.24&    [8.0-13.0]  &  1.01  \\
\hline
\hline
\bf{R~Vol}  &                               &   &   &    &  &    &       &       &    &    \\
\hline
Spectr  & \ldots            &          \ldots        & \ldots  &  \ldots &   \ldots    &   \ldots  &    \ldots       & \ldots   &  \ldots  \\
\hline
Photom  &  2800&3.85&	0.75& 390& -0.81& 1.69&  6&  2& 1.89&  [0.4-25.0] & \hphantom{0}1.08  \\
\hline
Interf  & 2800&3.85&	0.75& 390& -0.81& 1.69&  6&  2& 1.89 &  [8.0-13.0] &  23.40  \\ 
\hline
\hline
\bf{Y~Pav}  &                   &            &   &   &    &  &      &     &       &    &    \\
\hline
Spectr & \ldots            &          \ldots        & \ldots  &  \ldots &  \ldots  & \ldots &  \ldots   &  \ldots   &     \ldots       & \ldots   &  \ldots  \\
\hline 
Photom & 3200&	3.55&	0.75& 221&  -0.28	& 2.38&	  6&	  2& 0.36    & [0.4-25.0]&  11.15   \\
NO $\dot{M}$ & 2800&	4.00&	2.00& 525&  -0.53	& 2.38&	  4&	  2&  -    & [0.4-25.0]&  \hphantom{0}3.07  \\
\hline
Interf  &  3200&	3.55&	0.75& 221&  -0.28	& 2.38&	  6&	  2& 0.36   & [8.0-13.0]&  \hphantom{0}1.02    \\ 
\hline 
\hline
\bf{AQ~Sgr}  &                   &            &   &   &    &  &           &       &    &    \\
\hline
Spectr & \ldots            &          \ldots        & \ldots  &  \ldots &  \ldots     &     \ldots       & \ldots   &  \ldots  \\
\hline 
Photom  & 2600&	3.70&	0.75& 294&  -0.79	& 1.35&	  6&	  2&   1.69  & [0.4-25.0]&  \hphantom{0}1.41   \\
NO $\dot{M}$ & 2600&	3.85&	1.00& 390  &  -0.66	&  1.35&	  4&	  1&  -    & [0.4-25.0]&  \hphantom{0}1.01  \\
\hline
Interf &  2600&	3.70&	0.75& 294&  -0.79	& 1.35&	  6&	  2&   1.69   & [8.0-13.0]&  \hphantom{0}4.60  \\
\hline 
\hline
\bf{U~Hya}  &                   &            &   &   &    &  &   &        &       &    &    \\
\hline
Spectr & \ldots            &          \ldots        & \ldots  &  \ldots &  \ldots  & \ldots    &     \ldots       & \ldots   &  \ldots  \\
\hline 
Photom  & 3200 &	3.55&	0.75& 221&  -0.28	& 2.38&	  6&	  2&   0.36   & [0.4-25.0]&  13.58  \\
NO $\dot{M}$ & 2600&	3.85&	2.00&  390 &  -0.51	&  1.35&	  6&	  2&  -    & [0.4-25.0]&  \hphantom{0}3.20 \\
\hline
Interf  &  3200 &	3.55&	0.75& 221&  -0.28	& 2.38&	  6&	  2&   0.36   & [8.0-13.0]&  \hphantom{0}1.53 \\
\hline 
\hline
\bf{X~TrA}  &                   &            &   &   &    &  &           &       &    &    \\
\hline
Spectr  & \ldots            &          \ldots        & \ldots  &  \ldots &  \ldots  & \ldots &  \ldots    &     \ldots       & \ldots   &  \ldots  \\
\hline 
Photom & 2600&	4.00&	1.5&525& 	-0.79& 1.35&	  6&	  1& 2.51   & [0.4-25.0]&  14.70  \\
NO$R$,$I$ & 2600&	3.85&	2.0&390& 	-0.51&  1.35&	  6&	  1&  \ldots & [0.4-25.0]&  \hphantom{0}6.20  \\
NO $\dot{M}$ & 2600&	3.85&	2.00&  390 &  -0.51	&  1.35&	  6&	  1&  -    & [0.4-25.0]&  \hphantom{0}6.26 \\
\hline
Interf & 2600&	4.00&	1.5& 525 &	-0.79& 1.35&	  6&	  1&  2.51 &    [8.0-13.0]   &   \hphantom{0}1.04   \\
dist-20\% & 2600&	4.00&	1.5& 525&	-0.79& 1.35&	  6&	  1& 2.51  &  [8.0-13.0]   &   \hphantom{0}1.00  \\
\hline 
\hline

\end{tabular}\\
\end{table*}

\begin{figure}[!htbp]
\begin{center}
\resizebox{\hsize}{!}{
\includegraphics[width=0.7\textwidth, angle=90,bb=4 15 502 315]{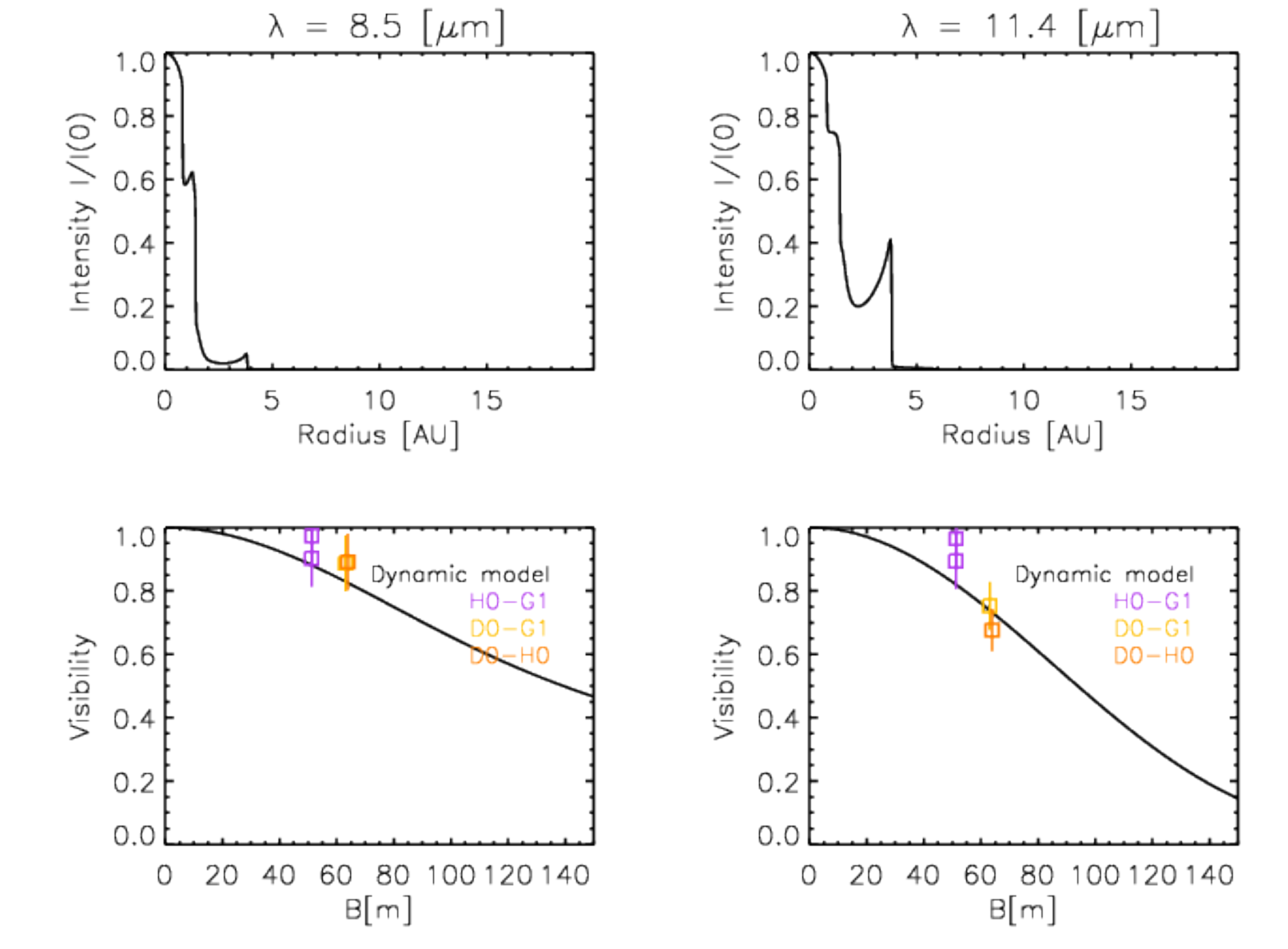}}
\caption{ Interferometric observational MIDI data of Y~Pav, in the case of the amount of SiC increased to $50$\% in the models, compared with the synthetic visibilities based on the DARWIN models. \textit{Up}: intensity profile at $11.3~\mu$m. \textit{Down}: visibility vs. baseline; the black line shows the dynamic model, the colored symbols illustrate the MIDI measurements at different baseline configurations.}
\label{y_pav_intens_11mu_50sic}
\end{center}
\end{figure}

 \begin{figure}[!htbp]
\begin{center}
\resizebox{\hsize}{!}{
\includegraphics[width=\textwidth, clip=true, angle=90,bb=251 7 502 327]{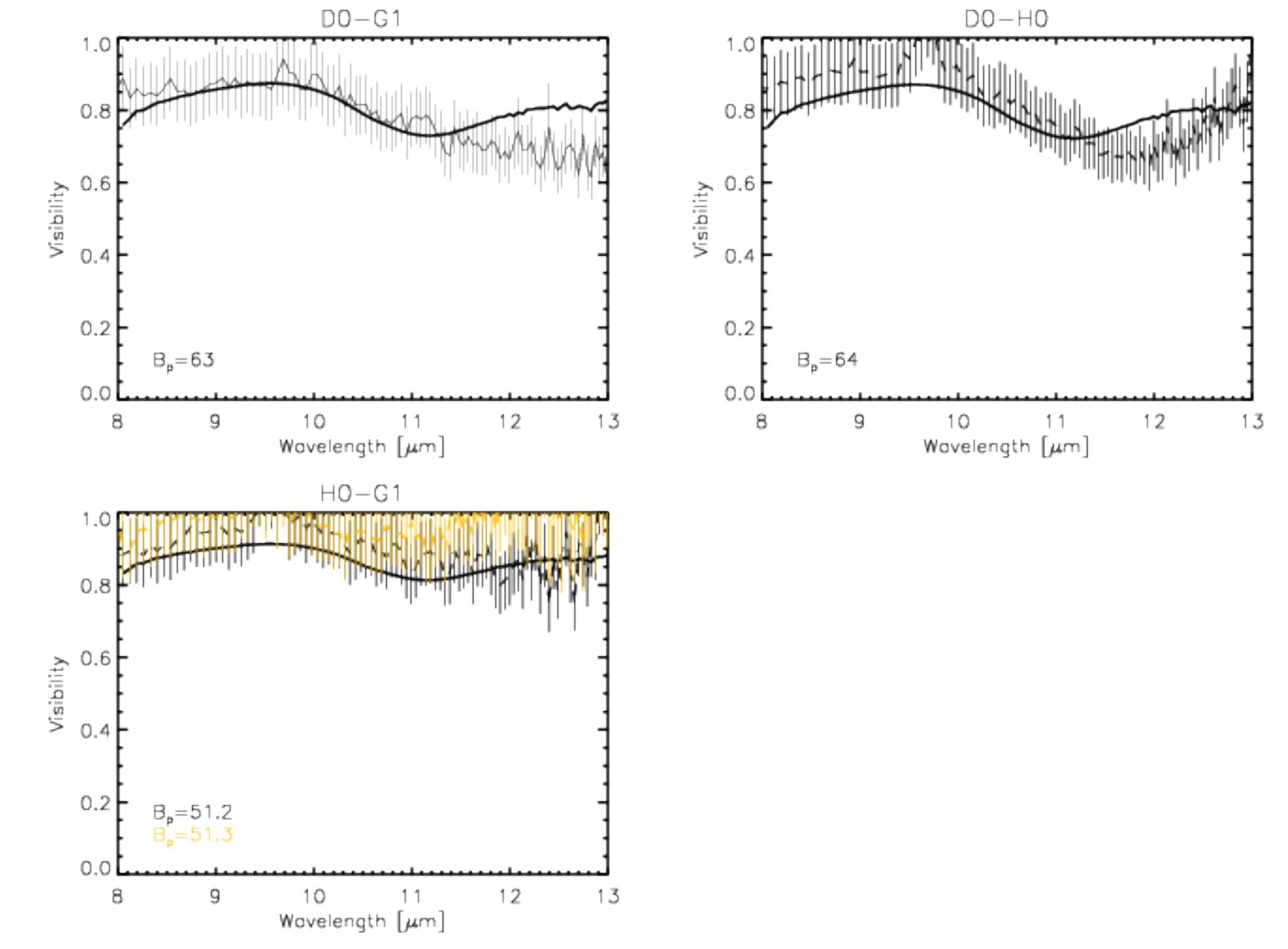}}
\caption{Y~Pav wavelength dependent visibilities in the MIDI range, for the only baseline configuration D0-H0, in the case of the amount of SiC increased to $50$\% in the models.}
\label{y_pav_wave_50percSiC}
\end{center}
\end{figure}

\subsubsection{Interferometric variability}\label{interfer-variability}
The data of Y~Pav, U~Hya and AQ~Sgr, have been taken at single epochs, therefore no interferometric variability can be assessed for those stars. The observations of R~Vol are one year apart, but very different in projected baseline ($B_\text{p}$) and projected angles (PA), a configuration that makes the variability check impossible to perform. The X~TrA observations are numerous and taken at the same time, but they are different in PA, thus the time-variability also can not be evaluated.

The only target for which a variability check could be performed is the Mira star R~Lep. 
For this check the two R Lep datasets at $B_\text{p} = 40$~m can be used. A small variation in the visibility level is noticeble between $9~\mu$m and $10~\mu$m (see Fig.~\ref{interf-mira}, panel (c)). The highest difference in visibility level is found at $9.7~\mu$m, where the variation of visibility is $\delta V = 0.072$ , which is barely significant compared to the typical errors of $\sim 10$~\% (see Fig.~\ref{y_pav_wave_50percSiC} for an example of the typical errors on the observed visibilities). The visibility level decreases when moving from pre-minimum ($\phi = 1.43$ at $B_\text{p} = 40$, PA$~=~147$) to post-minimum ($\phi = 0.66$ at $B_\text{p} = 40$, PA$~=~142$). This behavior goes in the same direction as the one found in the study of the Mira star V~Oph by \cite{ohnaka07}. Indeed they observed the star to be smaller close to the minimum of the visual phase, with a variation in the visibility level: $\delta V = 0.25$ between dataset \#$3$ (phase $\phi = 0.49$) and \#$6$ ($\phi = 0.69$) at $8.3$, $10.0$ and $12.5~\mu$m - see Fig.~2 in \citealp{ohnaka07}. The spatial frequency of the R~Lep measurements is smaller than that of the V~Oph data and gets even smaller when considering the smaller distance of V~Oph (237~pc, \cite{vanleeuwen07}). As can be seen from Fig.~2 of \cite{ohnaka07}, the variability decreases with decreasing spatial frequency. However, the lower visibilities found for R~Lep at all spatial frequencies indicate a significantly different structure when compared to V~Oph, probably in the sense of an overall larger extension of R~Lep. Thus a comparison of these two stars is difficult.

The above mentioned datasets do not only have a different variability phase but also belong to different cycles.  Since the projected baselines and projected angles are similar we used these data to check our approach of a combined fit of all data with individual times steps. An independent fit of those two observations to our models has been performed, leading to the same best fitting time-step. This result can be understood considering the temporal changes in the predicted visibilities.

Examples of these predicted changes in the visibilitie are shown in Figs.~\ref{rlep_range_bp34}, \ref{rlep_range_bp79} and \ref{rlep_range_visbase} for the best fitting model of R~Lep. For both Miras, the observed visibilities are at or close to the lower envelope of the predicted visibilities for all the time steps of the best fitting model. For the non-Miras the observations fall within the range of predicted visibilities and this range is larger than the one of the models for the Miras. The latter result is caused by cycle-to-cycle variation of the models, within each cycle the ranges are comparable. If our assumption of a small interferometric variability is correct, then the models for the Miras would predict a too large variability and would be too compact on average. If the real variability is not small we cannot draw any conclusion on the agreement between observed and predicted variations since the phase and $uv$-coverage of the MIDI observations is too small.  Concerning the wavelength dependence of the visibilities we note that the overall shape of the visibilities dispersed in wavelength is very similar from time step to time step, the major difference being in the overall level of visibility. This is important to keep in mind for the discussion in Sect.~\ref{visib-slope} and Sect.~\ref{SiC-dust}.

\section{Discussion}\label{discussion}

\subsection{The SEDs and visibilities}\label{sedandvisibilities}
Our attempts to reproduce the SED (photometry + IRTF spectrum in the R~Lep case) and interferometric MIDI data with DARWIN models show a strong improvement with respect to our previous study of RU~Vir. The models can reproduce the SEDs of all stars longward of $1~\mu$m quite well and also the visibilities between $8~\mu$m and $10~\mu$m.
In the Miras visibility vs. baselines profiles, the observations show a faster decline, and leveling off at longer baselines, in comparison to the non-Miras. This behavior is also predicted by the models which are significantly more extended for the Miras and have a more pronounced shell-like structure. This can be best seen by comparing R~Lep and Y~Pav (Fig.~\ref{interf-rlep-visbase} and Fig.~\ref{interf-ypav-visbase}). Indeed, since those two stars are located at very similar distances (see Table~\ref{table_starsparam}), the same baselines sample the same spatial frequency in AU$^{-1}$. It is also supported by the fact, that the best fitting models (with wind) for the non-Miras have a lower average mass loss rate and show only episodic mass loss. We remind the reader, that actually the best fitting models for the non-Miras were those without a wind but that we excluded those because of the known mass loss for these stars (Sect.~\ref{DARWIN models}). Both the windless and episodic models are characterized by rather compact atmospheres and weakly pronounced gas and dust shells.


In spite of these encouraging results in reproducing the observations, some notable (and partly systematic) differences remain. Therefore, our discussion will focus on three major parts: (1) differences at wavelengths shorter than $1~\mu$m; (2) differences in the visibilities longward of $10~\mu$m and (3) differences related to SiC dust. 

\subsubsection{Differences at wavelengths shorter than $1~\mu$m}\label{diff-1mu}
For all the stars in our sample, we noticed some differences at wavelengths shortwards of $1~\mu$m. In particular, the difference in the SEDs fit at the short wavelengths, appearing in Fig.~\ref{phot-mira}, and Fig.~\ref{phot-semi-irr}, could be caused by a possible combination of data related and model related effects. The data related ones are due to the stars variability, i.e.~lack of light curves, especially in $B$, $R$ and $I$ and partly in the IR. 

Concerning Semiregular and Irregular stars, their best fitting models are episodic models, as mentioned above, and thus show no regular light curve behavior. These two effects in combination introduce a larger uncertainty in the determination of mean magnitudes for the observations and models and are also responsible for the higher $\chi^2$ of the SED fits for non-Miras in comparison with the Miras. Deviations may also be due to the assumption of SPL in the models or uncertainties of the used data set for amC \citep{nanni16}.

\subsubsection{Differences in the visibilities longward of $10~\mu$m}\label{visib-slope}

Comparing the wavelength dependence of the visibilities for the Miras and the non-Miras with the models (see in Fig.~\ref{interf-mira} and Fig.~\ref{interf-semireg-irr}), and ignoring for the moment the differences in the SiC feature which will be discussed in the next section, one notices that at shorter baselines the Mira-models show an increase of visibility with wavelength which is not observed (full lines for the models, and the dashed lines for the observations respectively). A similar difference was also noticed by us for RU~Vir. In \citet{rau15} two explanations for this were discussed: (i)~a smoother density distribution than in the models and (ii)~a clumpy environment. A smoother density distribution with less pronounced dust shells seems possible as the models for the non-Miras do not show this slope in the wavelength-dependent visibilities, and these models generally have weakly pronounced shells (see Figs.~\ref{interf-ypav-visbase}, \ref{interf-uhya-visbase}, \ref{interf-aqsgr-visbase}, \ref{interf-xtra-visbase}). A clumpy environment is not excludable, but from our MIDI data we do not have any evidence of deviations from spherical symmetry, in particular all the differential phases are not significantly different from zero (see Paladini et al., in press). Furthermore, the slope of the wavelength-dependent model visibilities agrees quite well with the observations at the long baselines, where clumps should be more prominent. 

In this work we extended our search for other possible origins of the slope in the models. Using different opacity laws for amC (\citealp{zubko96} and \citealp{jaeger98}) did not change the slope. Also changing the distance within the expected uncertainties did not lead to a better agreement. 

The DARWIN models are known to poorly reproduce the SED around the $14~\mu$m feature of C$_2$H$_2$ and HCN, mostly due to uncertain opacity and chemistry data \citep{Gautschy-Loidl04}. 
We have checked the possible influence on the visibilities, by artificially removing the C$_2$H$_2$ and HCN contributions from the opacity, in the outer parts of the model. But also this experiment did not affect the slope. Thus, a smoother density distribution than the one produced by the models is the most likely explanation for the slope difference in the Mira models. 
Such a smoother density distribution could be caused by less pronounced shocks which might result from a different cooling function or different dust formation parameters. This has to be subject of further investigation.

Except for U~Hya, all the non-Miras show high visibility levels but no increase with wavelength. The differences to the models are partly related to SiC (see below). For Y~Pav they could be also due to calibration problems at the longest wavelengths (Paladini et al., in press). The high visibility levels also reduce the sensitivity to the differences in the models parameters, because at high visibility the difference between different time-steps, and different models becomes small. Indeed, the objects are only marginally resolved and therefore the observed MIDI data could not constrain the models very strongly.

\subsubsection{Differences related to SiC dust}\label{SiC-dust}

Lacking a consistent description of SiC formation in the models, the spectra and visibilities were calculated from the DARWIN models with the assumption that SiC condenses together with amorphous carbon (see Sect.~\ref{DARWIN models}). This means that the amount of SiC is proportional to the amount of amC grains and SPL is adopted. This assumption did not lead to major inconsistencies with the observations and is also not in disagreement with theoretical studies on SiC formation. These studies arrive at conflicting results for the condensation sequence of amC and SiC dust. \cite{gail} favor the scenario that amC dust condenses before SiC in the case of a stationary wind model. On the other hand, in the models of \cite{ferrarottiandgail06} SiC dust is the first dust component to start growing which is also supported by the work of \cite{cherchneff12}. Using models most comparable to our case, \cite{yasuda12} find that in the more likely case of non-LTE the formation region of the SiC grains is more internal and/or almost identical to that of the carbon grains, a scenario partially favored also by \cite{lagadec07}. A verification of this result requires the implementation of SiC condensation in the DARWIN models in a similar way as currently done for M-type stars \citep{hofner16}. This will be subject of future work.

In this context, we would like to underline that the visibilities level, lower \textit{in} the SiC feature than around this feature, has not to be interpreted as a larger extension of SiC with respect to amC and the molecular gas. This conclusion is only true for simple intensity profiles, while our stars have rather complex profiles and the contrast between the different shells containing dust and gas contributes to the influence on the level of visibility. This is illustrated in the comparison of the synthetic intensity and visibility profiles at $8.5~\mu$m and $11.4~\mu$m in Fig.~\ref{interf-rlep-visbase}. The lower visibility level around $11.3~\mu$m is solely due to the higher SiC opacity with respect to the one of amC.

Whenever the MIDI observations show a clear SiC dust feature, this $11.3~\mu$m SiC feature in the models is more peaked and narrower with respect to the observed one (see R~Lep and R~Vol in Fig.~\ref{interf-mira} and U~Hya in Fig.~\ref{interf-semireg-irr}). A similar effect was noted for RU~Vir, both for the spectra and the visibilities. For RU~Vir the spectral fit with hydrostatic models and More Of Dusty (MOD) could be improved by using the distribution of hollow speheres \citep{mod,rau15}. However, this distribution is not yet available for the DARWIN models, and thus it could not be tested.

Another free parameter for the fits is the fraction of Si condensed onto SiC. As explained in Sect.~\ref{DARWIN models}, we generally adopted a fraction of $10$~\%. Increasing this fraction to up to $50$~\% slightly improves the agreement for R~Lep. Also Y~Pav shows some improvements (see Fig.~\ref{y_pav_intens_11mu_50sic} and Fig.~\ref{y_pav_wave_50percSiC}), although for the latter star, the shape of the observed wavelength dependent visibilities is quite different and might not be due to SiC at all.  Given the still artificial treatment of SiC in the current models (see Sect~\ref{DARWIN models}), understanding the behaviour of the SiC feature in the model visibilities has to await a complete test of the SiC fraction for our stars has to await the above mentioned full implementation of SiC in the DARWIN models.

X~TrA is the only star for which no satisfactory fit of the wavelength dependent visibilities could be found. The shape of the model visibility vs. wavelength shows almost no SiC, while this is quite prominent in the data. Scaling the distance and increasing the SiC fraction did not remove this discrepancy. Also, we checked the two models closest in $\chi^2$ to the best fitting model (i.e. within $68$~\% of confidence level), in which the predicted SiC feature improves slightly, but the increase with wavelength is too steep, as in the case of Miras. The star is thus compact but apparently has a significant mass-loss. This combination cannot be reproduced by any of the models, and probably this is caused by the fact that the star is located in the parameter region of the models with episodic mass loss.

\subsection{Fundamental stellar parameters compared to literature, and evolutionary tracks}
	
The best fitting DARWIN models yield a number of parameters as listed in Tab.~\ref{tab_param_dyn}. 

For a comparison of temperature and luminosity with the literature values, we did not use the values given in the table as these refer to the hydrostatic initial model and thus not to the dynamic structure of the model at the time-step (i.e. phase) best fitting the interferometric data (see also \citealp{nowotny05}). Instead, we calculated for these time-steps a Rosseland diameter ($\theta_\text{Ross}$). The temperature of the time-step at this radius ($T_\text{Ross}$) is the corresponding effective temperature, i.e. the temperature at the Rosseland radius, defined by the distance from the center of the star to the layer at which the Rosseland optical depth equals $2/3$. From this and $\theta_\text{Ross}$, the luminosity $L_\text{Ross}$ is calculated. From the photometry of our stars we also derive the bolometric luminosity $L_\text{bol}$, a diameter $\theta_{\text{(V-K)}}$ and an effective temperature \textit{T}($\theta_{\text{(V-K)}}$) using the diameter/$\text{(V-K)}$ relation of \cite{vanbelle13}. The error on the luminosity is assumed to be about $40$~\%, on the basis of the distance uncertainty. The errors of the temperature are estimated through the standard propagation of error.
  
\begin{figure*}[!htbp]
\centering
\includegraphics[width=0.7\hsize, bb=0 0 504 684, angle=90]{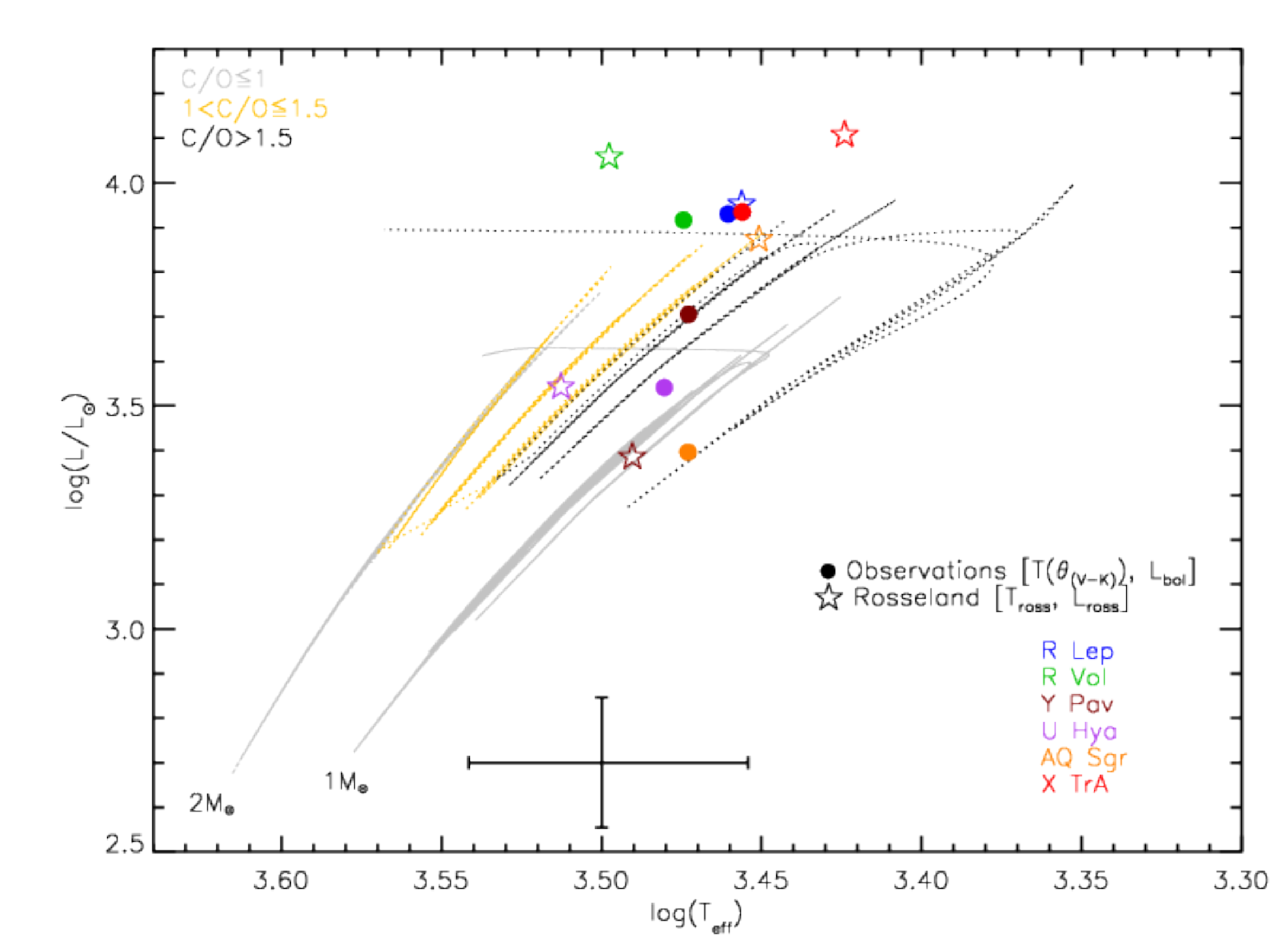}
\caption{\label{evolut-track} AGB region of the H-R diagram. The lines display solar metallicity evolutionary tracks from \cite{marigo13}: grey lines mark the regions of Oxygen-rich stars with $C/O<1.0$; yellow lines denote the region of C-rich stars with $1.0<C/O\leqslant~1.5$, while black lines with $C/O~>~1.5$. The numbers indicate the mass values at the beginning of the thermal pulsing (TP)-AGB. For better visibility, the track with $2~$M$_{\odot}$ is plotted with a dotted line. Different symbols and colors refer to the luminosity and effective temperature, estimated through the comparison in this work of the thef models with spectro-photometric-interferometric-observations. A typical error-size bar is shown in the lower side of the figure.}
\end{figure*}

The above various resulting stellar parameters are listed in Tab.~\ref{tab_theta} together with diameters at 8~$\mu$m and 12~$\mu$m from geometrical models (see Sect.~\ref{geom}). In Fig.~\ref{evolut-track} the temperatures and luminosities are compared to  thermally-pulsing (TP) AGB evolutionary tracks from \cite{marigo13}. Starting from the first thermal pulse, extracted from the PARSEC database of stellar tracks \citep{bressan12}, the TP-AGB phase is computed until all the envelope is removed by stellar winds. The TP-AGB sequences are selected with an initial scaled-solar chemical composition: the mass fraction of metals Z is $0.014$, and the one of helium Y is $0.273$.  In order to guarantee the full consistency of the envelope structure with the surface chemical abundances, that may significantly vary due to the third dredge-up episodes and hot-bottom burning, the TP-AGB tracks are based on numerical integrations of complete envelope models in which, for the first time, molecular chemistry and gas opacities are computed on-the-fly with the {\AE}SOPUS code \citep{marigoaringer09}. The results are shown in Fig.~\ref{evolut-track}, where the TP-AGB tracks for two choices of the initial mass on the TP-AGB, $M=1.0~$M$_{\odot}$, and $M=2$~M$_{\odot}$, are compared with the stars considered in this work.

We note the the TP-AGB model for $M=1.0$~M$_{\odot}$ does not experience the third dredge-up, hence remains with $C/O$<1 until the end of its evolution. Conversely, the model with $M=2$~M$_{\odot}$ suffers a few third dredge-up episodes which lead to reach $C/O$ > 1, thus causing the transition to the C-star domain. The location of the observed C-stars in the H-R diagram, as well as their $C/O$ ratios, appear to be nicely consistent with the part of the TP-AGB track that corresponds to the C-rich evolution. It is worth remarking that the current mass along the TP-AGB track is reduced during the last thermal pulses, which supports (within the uncertainties)  the relatively low values of the mass ($\sim 0.75-1.0~$M$_{\odot}$) assigned to some stars through the best fitting search on the DARWIN models datset.

Except for Y~Pav, the model luminosities and temperatures place the stars in the C-rich domain of the tracks. The model masses are all $1~$M$_{\odot}$ or less except for X~TrA. Such low current masses are in agreement with the tracks if the stars are in an advanced stage of the TP-AGB (when log$(L)~>~3.8$). This seems plausible for the Miras but not really the non-Miras. We note, that \cite{hinkle16} also found C-star masses of the order of $1~$M$_{\odot}$ from $1.5$~M$_{\odot}$. One should however keep in mind the uncertainties in the masses derived from the DARWIN models, and the ones predicted by the tracks.


The differences between the luminosity and temperature estimations derived from the models ($L_\text{Ross}$, $T_\text{Ross}$) and the observations ($L_\text{bol}$, \textit{T}($\theta_{\text{(V-K)}}$) are well within the error bars. Only for AQ~Sgr the difference in luminosity exceeds the error. This may be related to the above mentioned episodic mass-loss of the best fitting model. Literature values of luminosities can be found for three stars in \cite{mcdonald12} and they all agree within the uncertainties and considering the differences in the used data sets and the methods used. We underline the suprisingly good agreement between $T_\text{Ross}$ and the purely empirically determined \textit{T}($\theta_\text{(V-K)}$).

Temperature estimates in the literature are all based on fitting photometry with a combination of black bodies or spectra from hydrostatic model atmospheres and a dust envelope around it (\citealp{lorenz01}, \citealp{bergeat05}, \citealp{mcdonald12}). For each star different estimates typically differ by several hundred degrees and our values always are within the range of literature values. For R~Vol only one determination is found in the literature \citep{lorenz01} which gives a temperature $900$\,K lower than our $T_\text{Ross}$. This apparently large difference can be understood by the method used in \cite{lorenz01} which cannot take into account the very non-static character of a Mira variable and the strong radial overlap of photosphere and dusty envelope in C-rich atmospheres (for a detailed discussion on the concept of an effective temperature for these stars, see also Sect.~$3$~of~\citealp{nowotny05}). 

The diameters, $\theta_\text{Ross}$ and $\theta_{\text{(V-K)}}$ agree very well for the Miras, while the differences are larger for the non-Miras. This is probably again caused by the structure of models with episodic mass loss. Only R~Lep, U~Hya and AQ~Sgr have available observed $K$-diameters (see Table~\ref{tab_theta}). The values agree only roughly and there is no clear systematics in the differences between the three types of parameters. This can be understood by the fact that the three types of diameters sample quite different wavelength ranges and thus are affected quite differently by the non-hydrostatic atmospheric structure and the associated different molecular opacity contributions. 

The mass-loss values of the best fitting DARWIN models of the Miras are in reasonable agreement with the literature (see Table~\ref{table_starsparam}), while for the non-Miras we find large differences for AQ~Sgr and X~TrA. Again, the episodic mass loss of the models is the probable cause.

\begin{table*}[!htbp]
\centering
\small
\caption{\label{tab_theta} Observed and calculated temperatures and diameters.}
\begin{tabular}{llllllllllll}
\hline
\hline 
Target  &$\theta_\text{(V-K)}$~\textsuperscript{a}  &$\theta_\text{K}$& $\theta_{8}$ &$FWHM_{8}$ &$\theta_{12}$ & $FWHM_{12}$   &  $\theta_\text{{Ross}}$~\textsuperscript{e}  &  $T_\text{{Ross}}$ &  $L_\text{{Ross}}$ &  $T_{\theta_\text{(V-K)}}$ & $T_{\theta_\text{{K}}}$ \\
            &   [mas]               &   [mas]         &   [mas]               &   [mas]               &   [mas]      &   [mas]               &   [mas]            &  [K] & [L$_{\odot}$]  & [K] & [K] \\
\hline
R~Lep   &   $7.30$        &        $12.0\pm1.92$~\textsuperscript{b}     & 15.0 & 29.0$\pm$1.0    &12 &44.0$\pm$2.0     &  \hphantom{0}7.64 & $2860$ &  $8956$  & $2890 \pm 350$ &  $2250$  \\
R~Vol   &  3.60    &      \ldots  &$32.0 \pm 0.3$ &  \ldots  & $36.8 \pm 0.5$   & \ldots &   \hphantom{0}3.80  & $3140$ & $11438$  &   $2980 \pm 360$ & \ldots \\ 
Y~Pav   & 6.26  &  \ldots & \ldots & 5.3$\pm$1.1  &\ldots & 12.1$\pm$1.1 & \hphantom{0}4.00  & $3090$ & $2433$ & $2970 \pm 360$ & \ldots \\
U~Hya  &  9.62    &     $10.87\pm3.16$~\textsuperscript{c}    & $23.9\pm2.5$ &   \ldots &   $101.9 \pm \ldots$&   \ldots & \hphantom{0}8.30  & $3260$ & $3492$   & $3020 \pm 370$ & $2840$  \\ 
AQ~Sgr & 5.31    &        $6.13\pm0.52$~\textsuperscript{d} &     16.6$\pm$2.7 & \ldots      &32.9$\pm$2.8 & \ldots  &  10.18  & $2824$ & $7479$  &   $2970 \pm 360$ & $2770$ \\
X~TrA   & 9.78  &           \ldots     & 21.9$\pm$2.5  & \ldots   &39.0$\pm$3.0 & \ldots  & 13.82 & $2650$ & $12815$  &  $2860 \pm 350$ & \ldots \\
\hline
\hline
\end{tabular}\\
\textbf{Notes}. (a) Relation from \cite{vanbelle13}. (b) \cite{vanbelle97}.  (c) VINCI unpublished data. (d) \cite{richichi05}. (e) $\theta_\text{Ross}$ is the Rosseland diameter of the best fitting time-step of the corresponding best fitting model. 
\end{table*}

\section{Conclusions}\label{conclu}
In this work we presented a study on the atmospheres of a set of C-rich AGB stars, combining photometric and interferometric observations, comparing them consistently with a grid of dynamic model atmospheres.

Overall, we found that the fit of DARWIN models SEDs with the photometric and interferometric observations presented in this work show a strong improvement with respect of the one of RU~Vir. The best agreement is found for Mira stars, while for non-Miras the mass- loosing DARWIN models have more difficulties to reproduce the photometric observations and amplitudes at wavelengths shorter than $1~\mu$m. 

This could be related to the stars variability, since the photometric data in that wavelength region come from various studies, therefore a difference in phase is likely. With respect to our previous work on RU Vir, we notice a slight improvement in the agreement of the interferometric data with the models in terms of the level of the visibility vs. wavelength, but the difference in shape still remains and is probably due to the amount of condensed dust included in the models, as the experiments mentioned in Sect.~\ref{discussion} prove. Also, the observations show a consistency with the model assumption that SiC and amC condense together.

From our interferometric analysis, it resulted that the models for Miras appear to have a steeper slope in the visibility dispersed in wavelengths, with respect to the observed ones, and a larger extension, with respect to the models for non-Miras. The mass-loosing models for the non-Miras do not show this slope of increasing visibility with the wavelengths, generally have weakly pronounced shells and provide significantly better fits than the windless models.

Due to the sparse phase- and $uv$-coverage of the MIDI observations no conclusion can be drawn concerning the agreement between observed and predicted temporal variation in visibility.

We derived stellar parameters through the comparison of photometric and interferometric observations with dynamic models atmospheres and geometric models. Those parameters are summarized in Table~\ref{tab_theta} and Table~\ref{tab_param_dyn}. In the latter, errors on the temperatures are of the order of $\pm 400~$K and on the luminosity of the order of $2000~$L$_{\odot}$.

Models without the small particle limit assumption have lower condensation degrees, which probably implies less dust extinction in the visual region. Those models will represent a good test to verify the visual excess shown by some of the stars analysed in this study. Indeed, exclusion of the SPL assumption in a dust-shell changes the mid-IR interferometric shape and the temperature-structure. Indeed aready \cite{mattsson11} studied how, for certain cases, the effect of grain-size dependent opacities can be quite important, especially when strong dust-driven winds do not form in the SPL case, i.e.~for models near the limit of windless solutions, which might be of special relevance for the semiregular variables in our sample. Thus, models without the SPL assumption, compared with our observations, will be tested in a follow-up of this work (Rau et al. in prep). 
Another important aspect that is subject of ongoing studies is the development of mass-loss in mildly or irregularly pulsating stars (e.g. \citealp{liljegren16}). While the fits of the mass-loosing models for the Semi-regular and Irregular stars are quite reasonable for the SEDs and visibilities, the visual amplitudes cannot be reproduced. It is however interesting, that these models are all episodic or multi-periodic and thus do not have simple periodic light-curves as in the case of Miras.   

The second generation VLTI instrument MATISSE \citep{matisse06}, will allow imaging at the highest angular resolution. It will therefore be a perfect tool to better reconstruct the intensity profiles of the objects in this study, and to investigate the small scale asymmetries, in order to confirm or deny the asymmetric nature of the objects studied in this work. Also, MATISSE will also help to improve the variability study of those stars and the global distribution of molecules and dust. 

Additional interferometric observations of those targets will help us also to constrain the models better, e.g.~VLTI/PIONIER ($H$-band, \citealp{pionier11}), GRAVITY ($K$-band, \citealp{gravity08})  or Millimeter/submillimeter interferometric e.g.~ALMA measurements and VISIR observations could provide further constraints to solve the open questions.


\begin{acknowledgements}
We thank the anonymous referee for the constructive comments which helped to improve the quality of this paper.
This work is supported by the ``Abschlussstipendium fellowship'' of the University of Vienna, which GR thanks, and by the FWF project P23006. P.M. and B.A. acknowledge the support from the ERC Consolidator Grant funding scheme ({\em project STARKEY}, G.A. n.~615604). CP is supported by Belgian Fund for Scientific Research F.R.S.- FNRS. We thank the ESO Paranal team for supporting our VLTI/MIDI observations. We thank Thomas Lebzelter for the helpful discussions. This research has made use of the Jean-Marie Mariotti Center \texttt{Aspro} service \footnote{Available at http://www.jmmc.fr/aspro} and of SIMBAD database, operated at CDS, Strasbourg, France, and NASA/IPAC Infrared Science Archive. 
\end{acknowledgements}

\bibliography{mybib}{}
\bibliographystyle{aa}

\clearpage


\begin{appendices}


\section{Y~Pav and X~TrA models \textit{without} mass loss}\label{nomasslossypavxtra}

\begin{figure*}[!htbp]
\begin{center}
\resizebox{\hsize}{!}{
\includegraphics[width=0.7\textwidth, angle=90,bb=0 0 504 684]{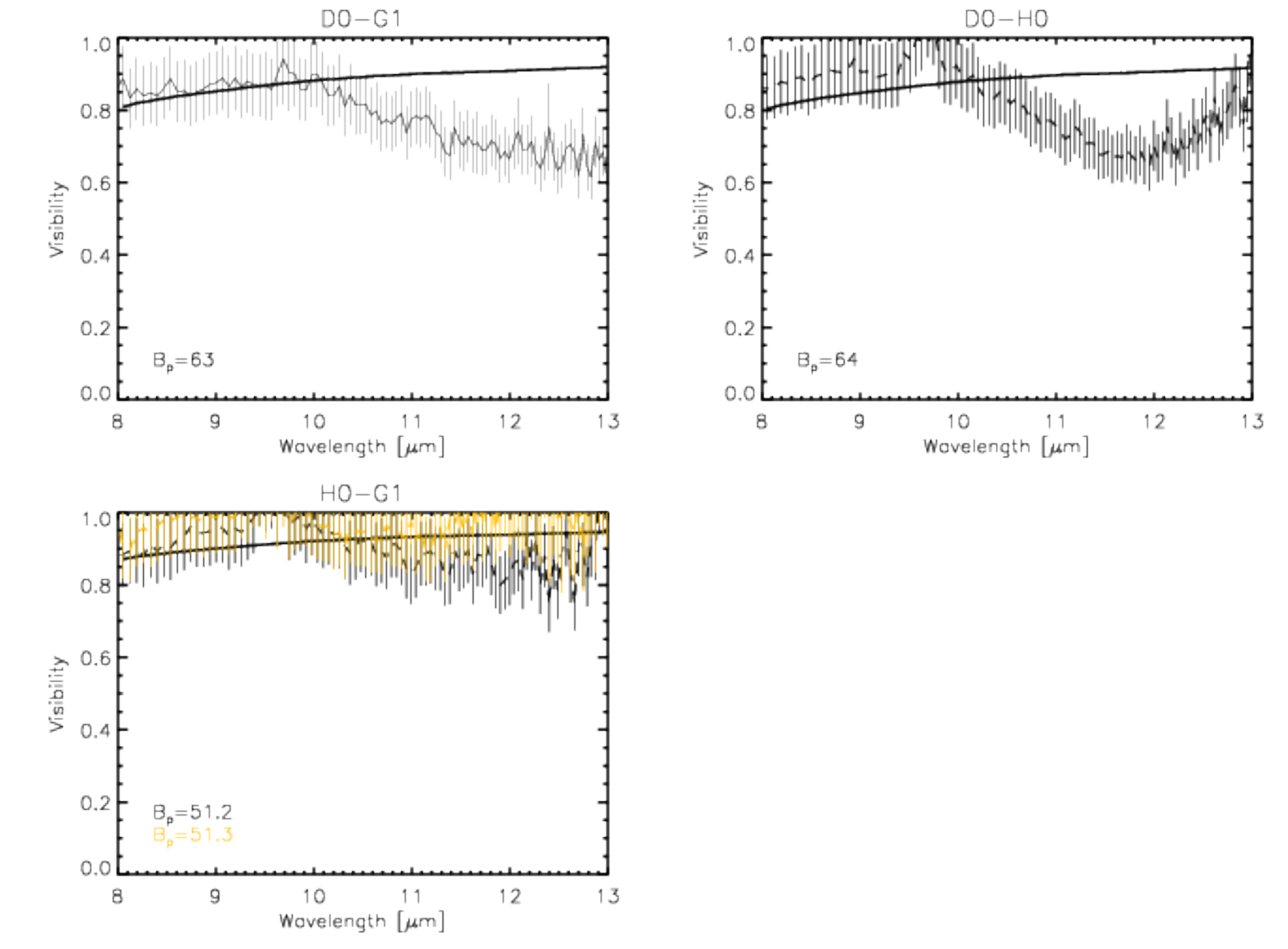}}
\caption{Observed visibility (dashed lines) dispersed over wavelengths of Y~Pav observations, compared to models (full line) \textit{without} mass loss. The different projected baselines are indicated in the color legend.}
\label{nomassloss-ypav}
\end{center}
\end{figure*}

\begin{figure*}[!htbp]
\begin{center}
\resizebox{\hsize}{!}{
\includegraphics[width=0.7\textwidth, angle=90,bb=0 0 504 684]{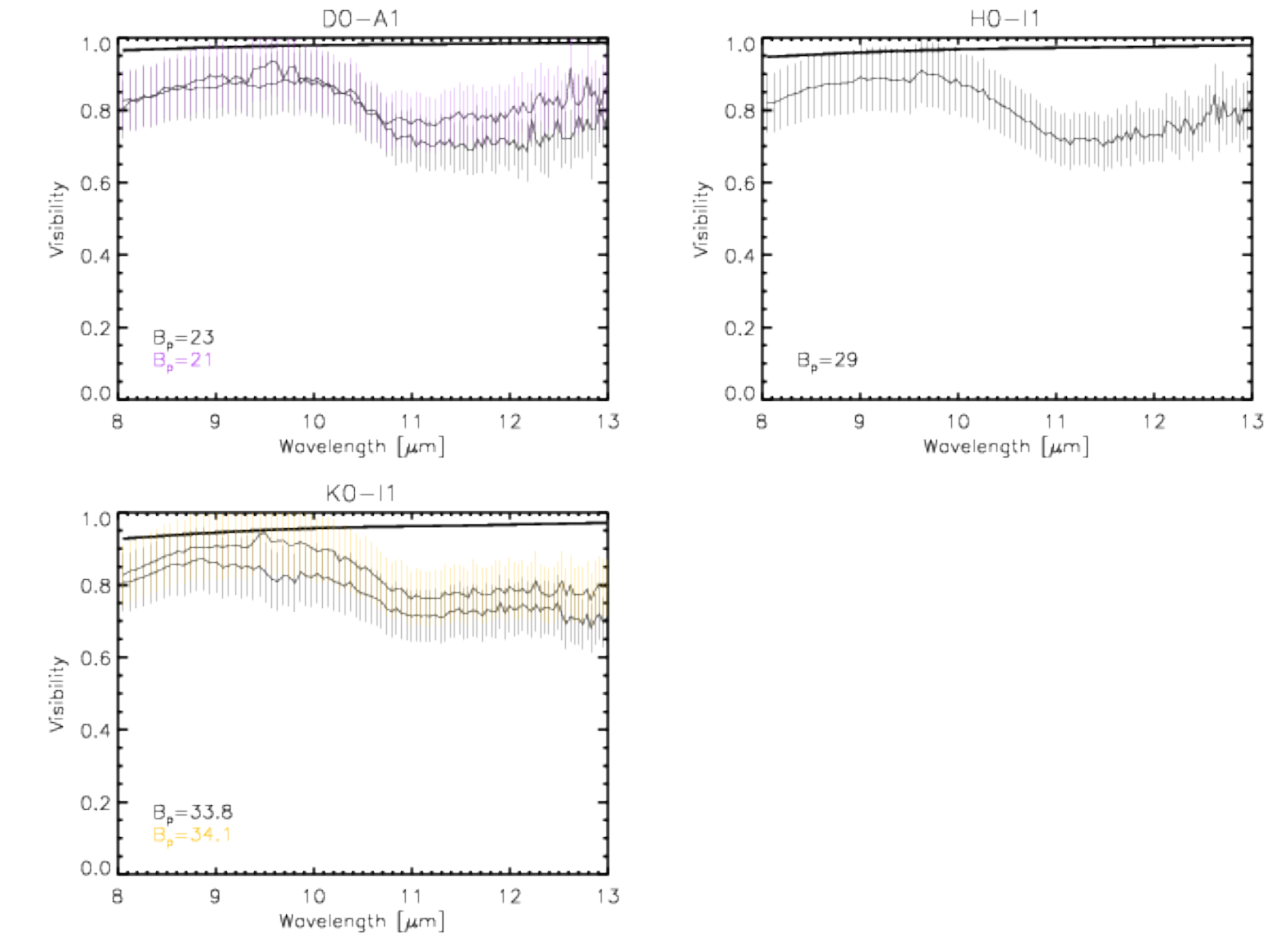}}
\caption{Observed visibility (dashed lines) dispersed over wavelengths of X~TrA observations, compared to models (full line) \textit{without} mass loss. The different projected baselines are indicated in the color legend.}
\label{nomassloss-xtra}
\end{center}
\end{figure*}

\section{Intensity profiles, visibilities vs. baselines, observing log}\label{observinglog}

\begin{table*}[!htbp]
\caption{\label{table_photometry} Photometric data from the literature. Different filters and different sources are given in units of mag. The consideration of the errors is described in Sect.~\ref{phot}.} 
\centering
\small
\begin{tabular}{lllllllllllllll}
\hline\hline             
Star  & $B$ & $V$ & $R$ & $I$ & $J$ & $H$ & $K$ & $L$ & $L'$ & $M$ & $N1$  & $N2$ & $N3$ & $IRAS12$ \\
     & [mag] & [mag] & [mag] & [mag] & [mag] & [mag] & [mag] & [mag] & [mag] & [mag] & [mag]  & [mag] & [mag] & [mag] \\
\hline
R~Lep & $11.71$ & \hphantom{0}$8.08$ & $5.84$ &  $4.58$ & $2.58$ & \hphantom{-}$1.17$ & \hphantom{-}$0.14$ & \hphantom{-}\ldots  & -$1.09$ & -$1.26$ & -$2.36$ & -$2.59$ & -$3.01$ & -$2.33$\\
R~Vol & $14.18$ & $10.68$ & $8.46$ &  $6.93$ & $5.08$ & \hphantom{-}$3.14$ & \hphantom{-}$1.71$ & \hphantom{-}\ldots & \hphantom{-}$0.08$  & -$0.61$ & \hphantom{-}\ldots & \hphantom{-}\ldots & \hphantom{-}\ldots & -$1.70$ \\
Y~Pav & \hphantom{0}$9.48$ & \hphantom{0}$6.28$ & $4.67$ &  $3.68$ & $1.76$ & \hphantom{-}$0.77$~\textsuperscript{a}  & \hphantom{-}$0.35$ & -$0.17$ & \hphantom{-}\ldots & \hphantom{-}$0.28$ & \hphantom{-}\ldots & \hphantom{-}\ldots & \hphantom{-}\ldots & -$0.56$ \\
AQ~Sgr & $10.40$ & \hphantom{0}$7.64$ & $5.67$ &  $4.43$ & $2.45$~\textsuperscript{a} & \hphantom{-}$1.31$~\textsuperscript{a} & \hphantom{-}$0.76$~\textsuperscript{a} &  \hphantom{-}$0.39$ & \hphantom{-}\ldots & \hphantom{-}$0.59$ & \hphantom{-}\ldots & \hphantom{-}\ldots & \hphantom{-}\ldots & -$0.29$ \\
U~Hya & \hphantom{0}$8.00$ & \hphantom{0}$5.03$ & $3.25$ &  $2.79$ & $0.89$ & -$0.25$~\textsuperscript{a} & -$0.59$ & -$0.91$ & \hphantom{-}\ldots & -$0.45$ & \hphantom{-}\ldots & \hphantom{-}\ldots & \hphantom{-}\ldots & -$1.69$ \\
X~TrA & \hphantom{0}$9.22$ & \hphantom{0}$5.71$ & $4.59$ &  $5.51$ & $1.09$ & -$0.01$~\textsuperscript{a} & -$0.59$ & -$0.95$ & \hphantom{-}\ldots & -$0.45$ & \hphantom{-}\ldots & \hphantom{-}\ldots & \hphantom{-}\ldots & -$1.67$ \\
\hline
\hline
\end{tabular}\\
\textbf{Notes}. (a): $2MASS$ photometry. 
\end{table*}

 \begin{figure*}[!htbp]
\begin{center}
\resizebox{\hsize}{!}{
\includegraphics[width=\textwidth, bb=0 0 504 684, angle=90]{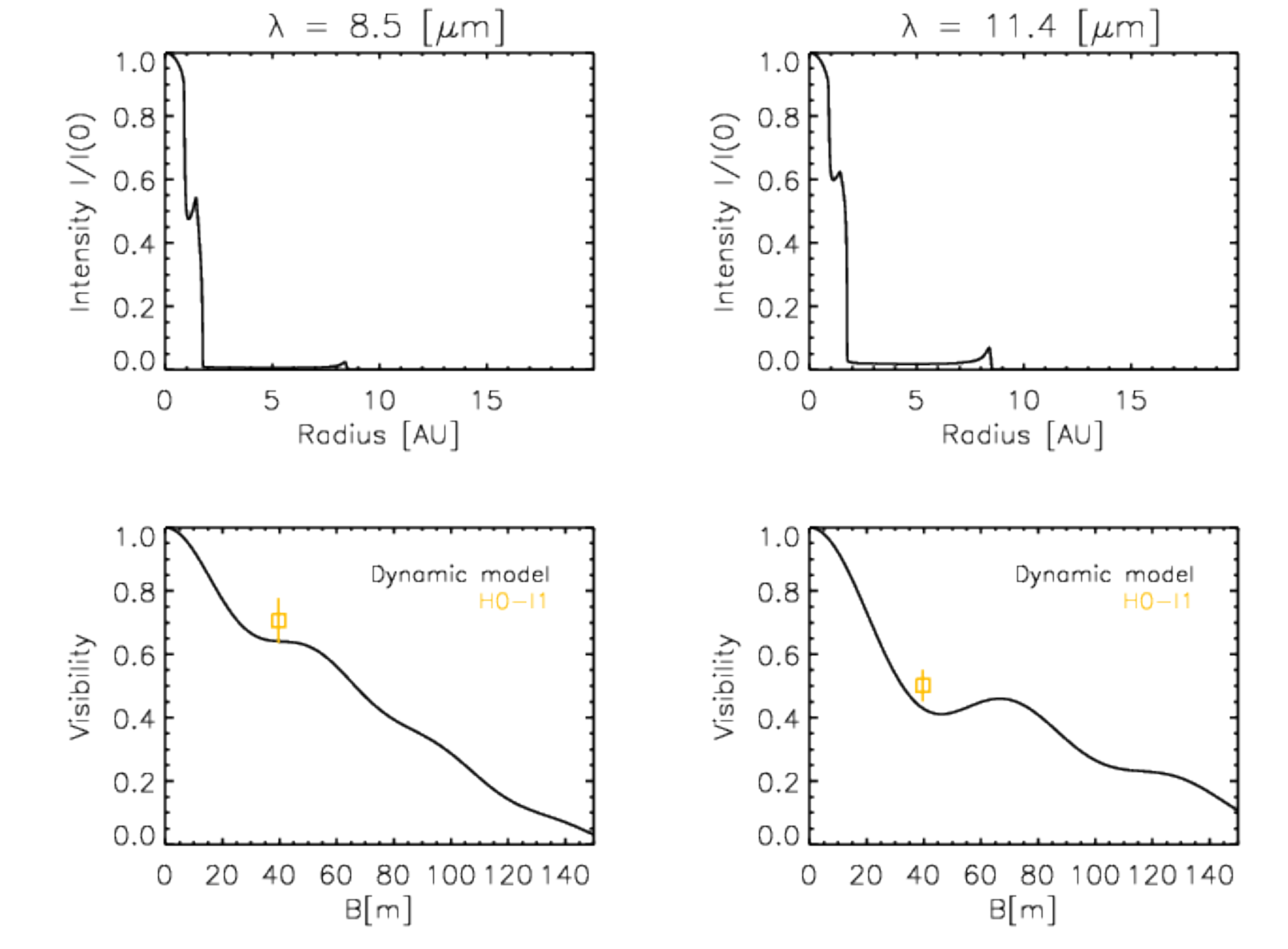}}
\caption{Interferometric observational MIDI data of U~Hya, compared with the synthetic visibilities based on the DARWIN models. \textbf{Up}: intensity profile at two different wavelengths: $8.5~\mu$m and $11.4~\mu$m. \textbf{Down}: visibility vs. baseline; the black line shows the dynamic model, the colored symbols illustrate the MIDI measurements at different baselines configurations.}
\label{interf-uhya-visbase}
\end{center}
\end{figure*}

   	\begin{figure*}[!htbp]
\begin{center}
\resizebox{\hsize}{!}{
\includegraphics[width=\textwidth, bb=0 0 504 684, angle=90]{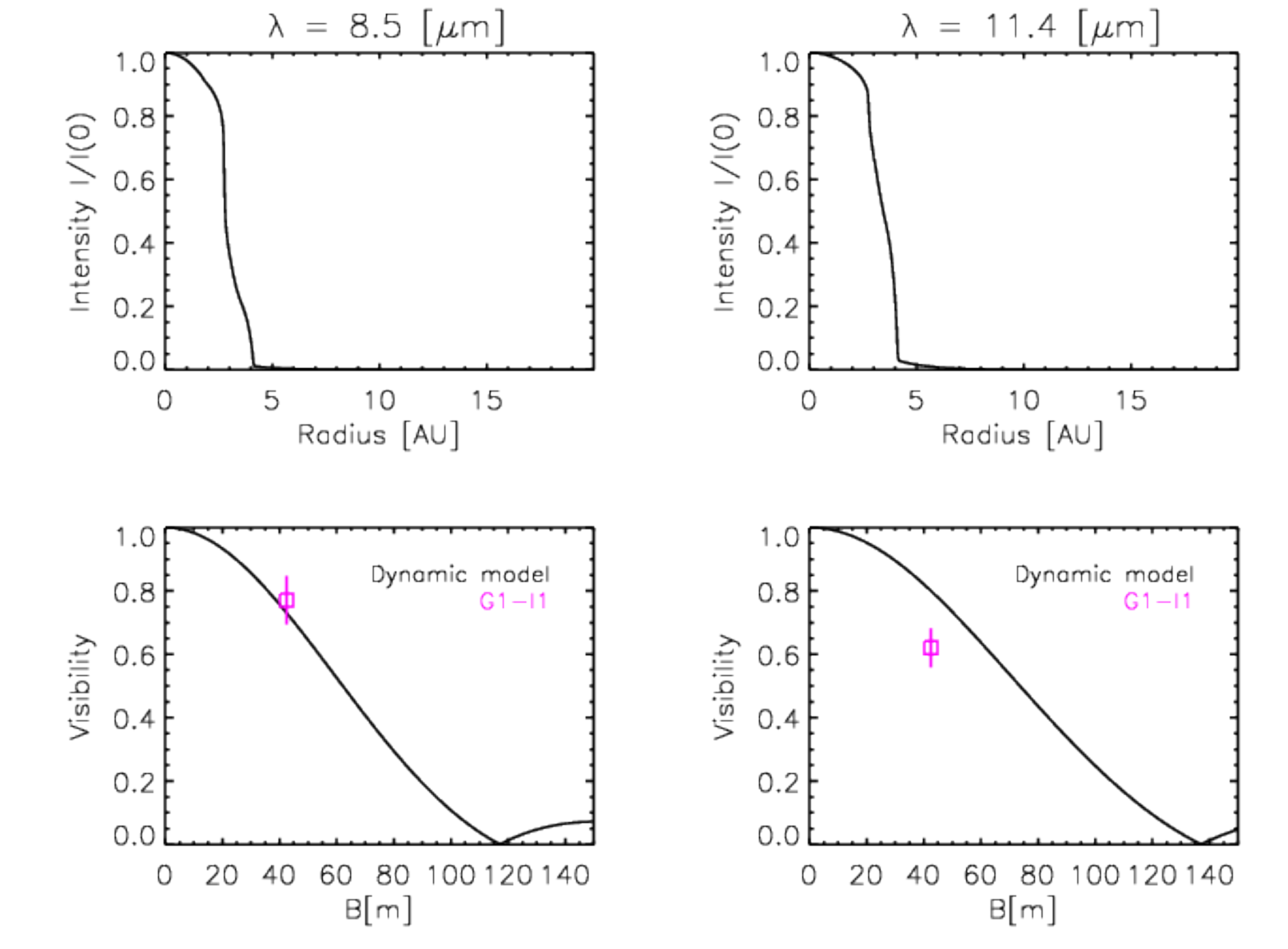}}
\caption{Same as Fig.~\ref{interf-uhya-visbase}, but for AQ~Sgr.}
\label{interf-aqsgr-visbase}
\end{center}
\end{figure*}

\begin{figure*}[!htbp]
\begin{center}
\resizebox{\hsize}{!}{
\includegraphics[width=\textwidth, bb=0 0 504 684, angle=90]{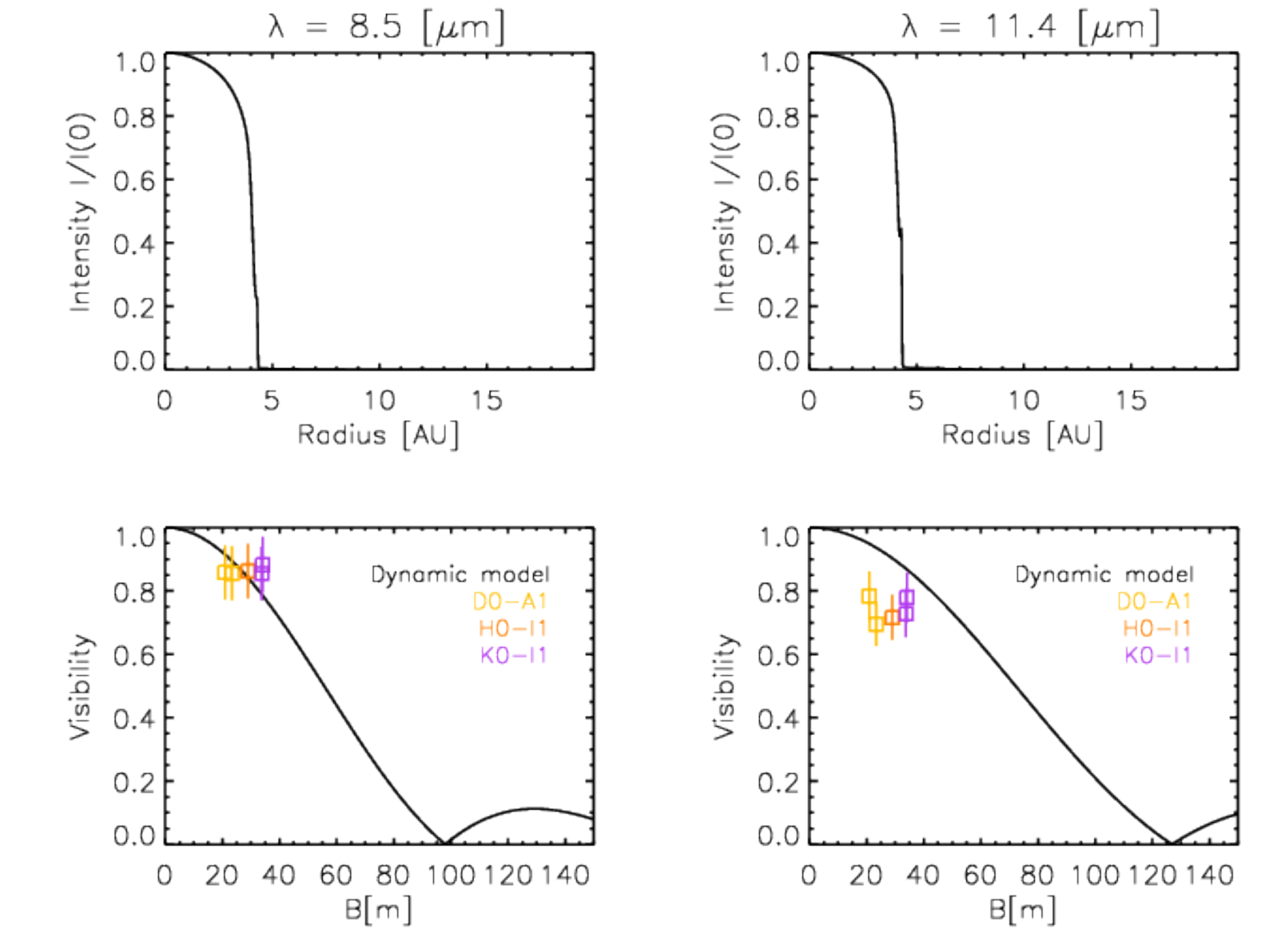}}
\caption{Same as Fig.~\ref{interf-uhya-visbase}, but for X~TrA.}
\label{interf-xtra-visbase}
\end{center}
\end{figure*}

\begin{table*}
\caption{\label{tab_midi_observ_r_vol} Journal of the MIDI observations of R~Vol.} 
\centering
\begin{tabular}{lllllllll}
\hline\hline
Target & UT date \& time & Config. & $B_\text{p}$ & PA & Seeing & Airmass &   Mode & Phase  \\
& & & [m] & [$^\circ$]&[''] &\\
\hline\hline

R~Vol & 2012-10-05 T08:42:06   &        G1-A1  & \hphantom{0}74.2 &
         -37     & 0.58 &
        1.59     &     SCI-PHOT & 0.06 \\
HD 32887 &  	2012-10-05 T08:11:05  &
        \ldots &
        \hphantom{0}79.2 &  -106 &
        0.49 &  1.01 &   \ldots & \ldots \\
HD 82668 & 2012-10-05 T08:28:26 	 &
        \ldots &
        127.4 &  -85 &
        0.66 &  1.94 &   \ldots & \ldots \\

\hline
       R~Vol & 2013-01-18 T02:09:01   &        A1-K0  & 126.0 &
         -32     & 0.58 &
        1.57    &     SCI-PHOT & 0.29  \\
HD 82668 & 2013-01-18 T01:54:03 	 &
        \ldots &
        \hphantom{0}40.7 &  -134 &
        0.49 &  1.81 &   \ldots & \ldots \\
\hline
\hline
\end{tabular}
\end{table*}

\begin{table*}
\caption{\label{tab_midi_observ_u_hya} Journal of the MIDI observations of U~Hya.} 
\centering
\begin{tabular}{lllllllll}
\hline\hline
Target & UT date \& time & Config. & $B_\text{p}$ & PA & Seeing & Airmass &   Mode & Phase  \\
& & & [m] & [$^\circ$]&[''] &\\
\hline\hline

U~Hya & 2011-03-11 T01:24:49   &        H0-I1  & 39.70 &
         -112.6     & 0.85 &
        1.29     &     SCI-PHOT & 0.85 \\
HD 81797 &  2011-03-11 T01:09:56.015	 &
        \ldots &
        40.76 &  -125.6 &
        0.85 &  1.42 &   \ldots & \ldots \\
HD 81797 & 2011-03-11 T01:41:04	 &
        \ldots &
        40.70 &  -134.2 &
        0.82 &  1.43 &   \ldots & \ldots \\

\hline
\hline
\end{tabular}
\end{table*}

   

 \begin{figure*}[!htbp]
\begin{center}
\resizebox{\hsize}{!}{
\includegraphics[width=0.5\textwidth, angle=0, bb=0 0 504 360]{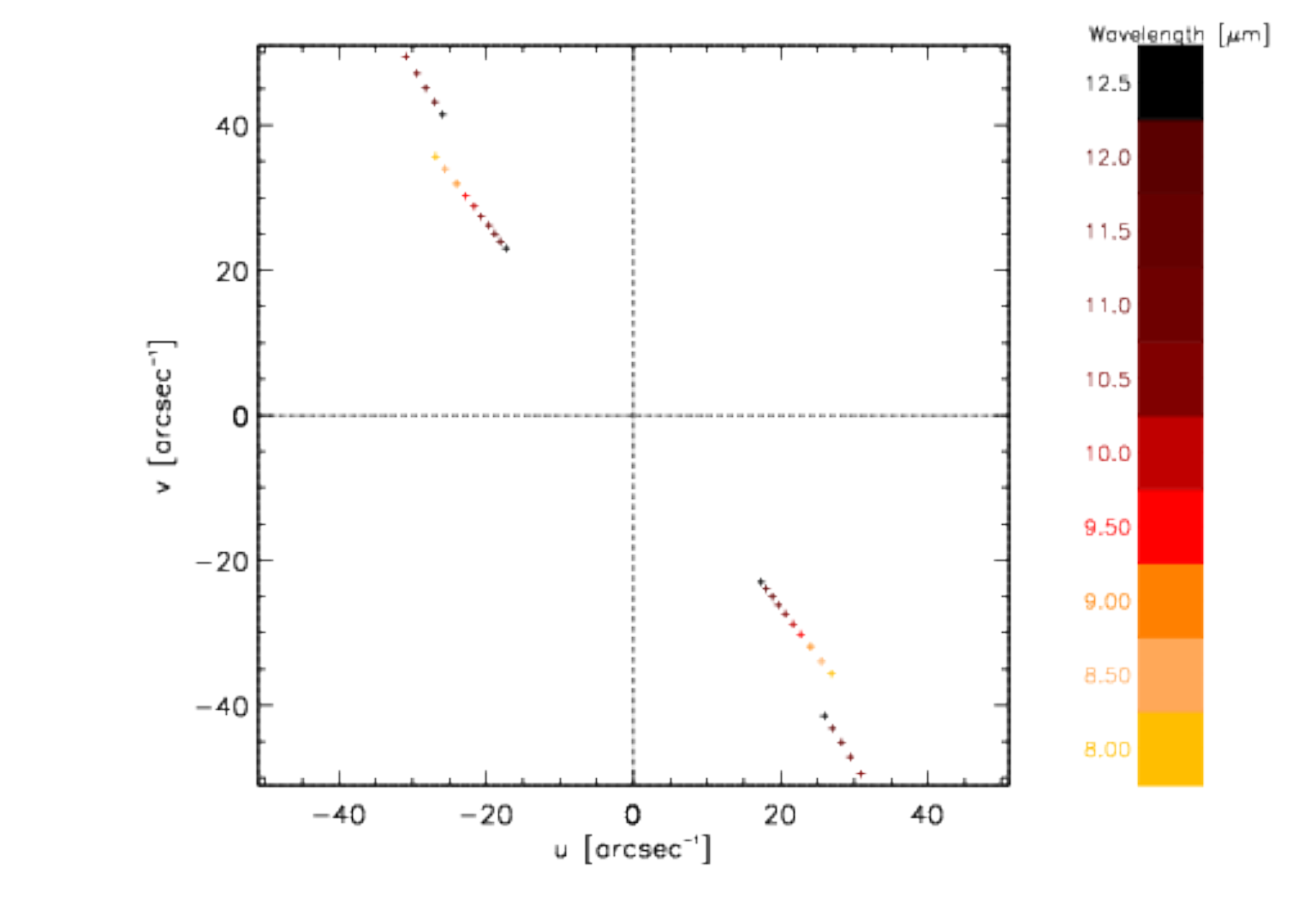}
\includegraphics[width=0.5\textwidth, angle=0, bb=0 0 504 360]{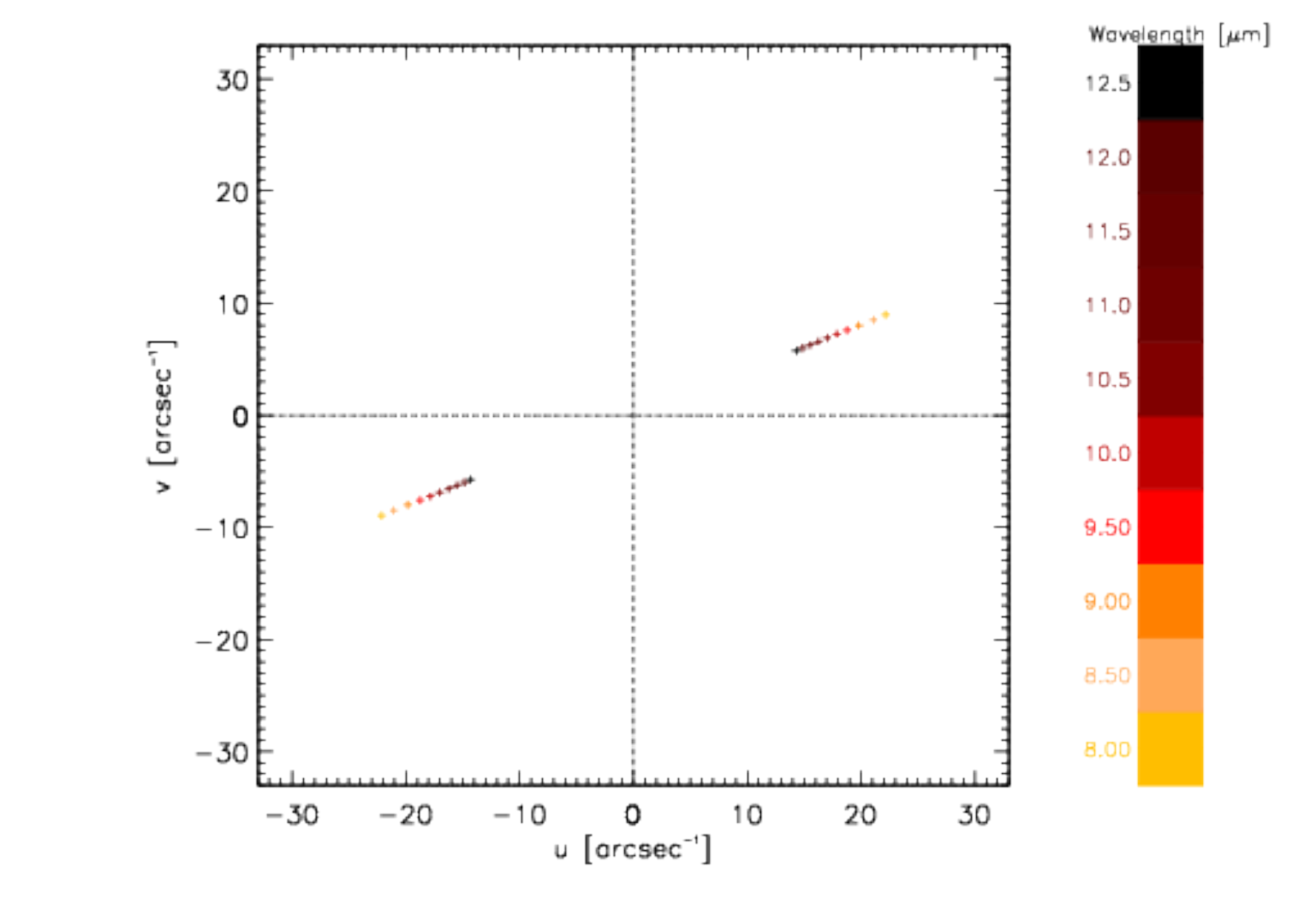}}
\caption{$uv$-coverage of the MIDI observations of R~Vol (left side) and U~Hya (right side) listed in Table~\ref{tab_midi_observ_r_vol} and \ref{tab_midi_observ_u_hya}, dispersed in wavelengths.}
\label{uv-coverage}
\end{center}
\end{figure*}

\end{appendices}

\end{document}